\documentclass[aps,a4,longbibliography,notitlepage]{revtex4-1}
\usepackage[latin9]{inputenc}
\usepackage{array}
\usepackage{textcomp}
\usepackage{multirow}
\usepackage{amssymb}
\usepackage{graphicx}

\makeatletter

\providecommand{\tabularnewline}{\\}

\usepackage{aas_macros}

\usepackage{ifpdf}\ifpdf 
    \usepackage[pdftex,     % sets up hyperref to use pdftex driver
            plainpages=false,   % allows page i and 1 to exist in the same document
            breaklinks=true,    % link texts can be broken at the end of line
            colorlinks=true,
            pdftitle=My Document
            pdfauthor=My Good Self
           ]{hyperref} 
    \usepackage{thumbpdf}
\else 
    \usepackage{graphicx}       % to include graphics
    \usepackage{hyperref}       % to simplify the use of \href
\fi 
\makeatother

\begin{document}

\title{A cosmological exclusion plot:\\
Towards model-independent constraints on modified gravity \\
from current and future growth rate data}

\author{Laura Taddei$^{1,2}$, Luca Amendola$^{2}$}

\affiliation{$^{1}$Dipartimento di Fisica e Scienze della Terra, Università di
Parma, Viale Usberti 7/A, I-43100 Parma, Italy. }

\affiliation{$^{2}$Institut für Theoretische Physik, Ruprecht-Karls-Universität
Heidelberg, Philosophenweg 16, 69120 Heidelberg, Germany.}
\begin{abstract}
Most cosmological constraints on modified gravity are obtained assuming
that the cosmic evolution was standard $\Lambda$CDM in the past and
that the present matter density and power spectrum normalization are
the same as in a $\Lambda$CDM model. Here we examine how the constraints
change when these assumptions are lifted. We focus in particular on
the parameter $Y$ (also called $G_{\mathrm{eff}}$) that quantifies
the deviation from the Poisson equation. This parameter can be estimated
by comparing with the model-independent growth rate quantity $f\sigma_{8}(z)$
obtained through redshift distortions. We reduce the model dependency
in evaluating $Y$ by marginalizing over $\sigma_{8}$ and over the
initial conditions, and by absorbing the degenerate parameter $\Omega_{m,0}$
into $Y$. We use all currently available values of $f\sigma_{8}(z)$.
We find that the combination $\hat{Y}=Y\Omega_{m,0}$, assumed constant
in the observed redshift range, can be constrained only very weakly
by current data, $\hat{Y}=0.28_{-0.23}^{+0.35}$ at 68\% c.l. We also
forecast the precision of a future estimation of $\hat{Y}$ in a Euclid-like
redshift survey. We find that the future constraints will reduce substantially
the uncertainty, $\hat{Y}=0.30_{-0.09}^{+0.08}$ , at 68\% c.l., but
the relative error on $\hat{Y}$ around the fiducial remains quite
high, of the order of 30\%. The main reason for these weak constraints
is that $\hat{Y}$ is strongly degenerate with the initial conditions,
so that large or small values of $\hat{Y}$ are compensated by choosing
non-standard initial values of the derivative of the matter density
contrast. 

Finally, we produce a forecast of a cosmological exclusion plot on
the Yukawa strength and range parameters, which complements similar
plots on laboratory scales but explores scales and epochs reachable
only with large-scale galaxy surveys. We find that future data can
constrain the Yukawa strength to within 3\% of the Newtonian one if
the range is around a few Megaparsecs. In the particular case of $f(R)$
models, we find that the Yukawa range will be constrained to be larger
than $80$ Mpc$/h$ or smaller than 2 Mpc$/h$ (95\% c.l.), regardless
of the specific $f(R)$ model.
\end{abstract}
\maketitle

\section{Introduction}

Testing possible modifications of gravity at very large scales is
currently one of the most interesting research activity in cosmology.
Modifications of standard gravity are often modeled by introducing
one or more additional mediating fields in the gravitational Lagrangian.
One of the most well studied example is the so-called Horndeski theory,
which adds to the Einstein-Hilbert Lagrangian a single scalar field
that obey the most general second order equation of motion \cite{Horndeski:1974}.

As shown in several papers (e.g. \cite{2008JCAP...04..013A,DeFelice:2011hq,Silvestri:2013ne,2012JCAP...06..032Z}),
a generic modification of gravity introduces at linear perturbation
level two new functions that depend only on background time-dependent
quantities and, in Fourier space, on the wavenumber $k$ . One function,
that we denote here with $Y(t,k)$ (sometimes also called $G_{\mathrm{eff}}$),
modifies the standard Poisson equation, while the second one, $\eta(t,k)$,
the anisotropic stress or tilt, provides the relation between the
two gravity potentials $\Psi,\Phi$. In standard gravity, one has
$Y=\eta=1$.

In the so-called quasi-static regime (i.e. for linear scales that
are below the sound horizon) of the Horndeski models, and also in
some cases \cite{Konnig:2014xva} of bimetric models \cite{Hassan:2011hr},
the two functions $Y,\eta$ take a particularly simple form and can
be directly constrained through observations \cite{horn,2013PhRvD..87b3501A}.
In particular, one can use observations of weak lensing, redshift
distortions and galaxy clustering to constrain or detect modifications
of gravity at cosmological scales \cite{Amendola:2013qna}.

One problem of these techniques is that often one makes explicitly
or implicitly several assumptions that might not be warranted by current
data. For instance, one often assumes that the behavior of the cosmological
model before dark energy domination, i.e. essentially at any time
except very recently, is the standard radiation and matter dominated
universe. While we have at least some proof that the radiation epoch
had to be close to standard, otherwise one would see deviations from
the standard big bang nucleosynthesis and on the microwave background
sky, we have much less robust data concerning the matter dominated
era, in particular between decoupling and now. For instance, models
in which the dark energy was a substantial fraction of the cosmic
energy at high redshift \cite{EDE2,Pettorino_etal_2013} cannot yet
be excluded.

We identify in particular three assumptions that are very commonly
made (at least one of them is included in, for instance, \cite{2008PhRvD..78l3010G,Narikawa_Yamamoto_2010,Hirano:2011wj,Shi:2012ci,Macaulay:2013swa,Samushia:2013yga,Arkhipova:2014dha,2014JCAP...05..042S})
and which are certainly acceptable in some cases but that, in reality,
are not necessariy warranted in more general gravity theories. First,
we do not know what is the present value of the matter density fraction
$\Omega_{m,0}$. If we take it from distance measurements (supernovae,
baryon acoustic oscillation) then one should be aware that the observed
quantity is the expansion rate $H(z)$ and not the equation of state
$w(z)$ or $\Omega_{m,0}$. In fact, the EOS $w(z)$ depends on assuming
a value of $\Omega_{m,0}$, and viceversa \cite{Kunz:2006ca}. Of
course if $w(z)$ is parametrized by a small number of parameters
then one can get also $\Omega_{m,0}$ from the distance data, but
the estimation will depend on the chosen parametrization. Moreover,
$\Omega_{m,0}$ cannot be determined without ambiguity with other
techniques, e.g. from weak lensing (e.g. \cite{More:2014uva}) or
$X$-ray temperature in clusters (e.g. \cite{2013MNRAS.432..973R}),
since these estimates always assume standard Newtonian gravity.

Second, we do not know what is the present value of the power spectrum
amplitude $\sigma_{8}$. In fact, any estimate of $\sigma_{8}$, through
e.g. weak lensing (see e.g. \cite{More:2014uva}), cosmic microwave
background (e.g. \cite{Planck_016}), or cluster abundances (see e.g.
\cite{2013MNRAS.432..973R}, \cite{PlanckSZ13}), depends again on
assuming a particular (normally, Newtonian), theory of gravity.

Third, when we obtain the theoretical behavior of the linear perturbation,
by integrating the matter conservation equations, we need to assume
some initial condition for the matter density contrast $\delta_{m}$
and the peculiar velocity divergence $\theta_{m}$ (or equivalently
on $\delta_{m}$ and $\delta_{m}'$, the prime being from now on the
derivative with respect to the e-folding time $N=\log a$). Typically,
this problem is bypassed assuming that the evolution in the past (say,
for redshifts $z\gg1$ ) was identical to a matter dominated universe
so that $\delta_{m}\sim a$ and $\delta_{in}'=\delta_{in}$ (of course
since we are in the linear regime one can always choose freely one
of the two inital conditions, say $\delta_{in}$). However, if we
do not know the cosmological model in the past, we cannot fix $\delta'_{in}$.
For instance, in some coupled dark matter-dark energy model the perturbations
grow faster than in $\Lambda$CDM during the matter epoch due to the
fact that the dark energy field is not negligible (e.g. \cite{2001PhRvD..64d3509A});
in this case $\delta'_{in}>\delta_{in}$. Similarly, in a Brans-Dicke
model with coupling $\omega$ one has $\delta'_{in}=(2+\omega)\delta_{in}/(1+\omega)$
\cite{1969PThPh..42..544N}; although $\omega$ has to be very large
to pass local gravity constraints, if a screening mechanism is present
these bounds becomes very weak.

In this paper we wish to examine what constraints one can still get
on modified gravity, in particular on $Y$, when all three assumptions,
on $\Omega_{m,0},\sigma_{8}$ and $\delta'_{in}$, are lifted by marginalizing
over all the non-degenerate parameters. We will consider both current
data and forecasted data from a future experiment that approximates
the Euclid%
\footnote{http://www.euclid-ec.org%
} survey \cite{Euclid-r}. We call this a model-independent approach,
although of course we are still making several model-dependent assumptions,
like for instance that we are really dealing with linear scales in
the sub-horizon regime and that matter is conserved. We also assume
for simplicity that matter is a pressureless fluid and that the background
is well approximated by a $\Lambda$CDM behavior during the redshift
range that we consider, although both these assumptions can be easily
generalized. One has also to bear in mind that it is possible to modify
gravity leaving the function $Y$ unaltered (but not $\eta$, when
properly defined in the Jordan frame, see discussion in \cite{Saltas:2014dha})
so that even finding $Y=1$ does not guarantee Einsteinian gravity.

We use the $f\sigma_{8}(z)$ data obtained through the redshift distortion
method \cite{Percival:2008sh} and collected in \cite{Macaulay:2013swa,More:2014uva}.
This method does not rely on assuming standard gravity, contrary to
methods based, for instance, on extrapolation from CMB data, on weak
lensing, cluster abundances, or galaxy power spectra.

The conclusion is that both present and future data have little chance
to set stringent constraints on $Y$ if one wants to be as much model-independent
as possible. We find in particular a strong degeneracy between the
initial conditions and $Y$ that allows both relatively large and
small values of $Y$. It is important to remark that we simplified
out task by setting $Y$ constant in time (the space dependence is
either ignored as well or introduced according to the Horndeski model,
see below). If we include an arbitrary time dependence, then the constraints
would evaporate completely for current data, since we would have one
datum at each redshift and one free parameter per redshift \emph{plus}
initial conditions and $\sigma_{8}$. For future data, where one can
in principle obtain several data points at different $k$'s for each
redshift, the constraints would not disappear but weaken a lot, and
even more so if the initial conditions are taken to be $k$-dependent.
Clearly, obtaining any constraint at all would be totally impossible
if the $Y$ function, instead of being restricted to follow the Horndeski
form, were a completely arbitrary function of time and space.

This leaves one only two escape routes to obtain stronger constraints
on modified gravity through cosmological observations at linear scales.
The first one is to forget model-independency and assume specific
modified gravity models. Then one can estimate the model-specific
$\sigma_{8},\Omega_{m,0}$ and the initial conditions and confine
$Y$ within a much narrower region. The second one is to use a different
parameter to test modified gravity. In Ref. \cite{Amendola:2012ky}
it has been shown that the anisotropic parameter $\eta$ is a useful
probe of gravity since it is independent of $\sigma_{8}$ and it can
be estimated from observations through an algebraic relation, i.e.
without the need of choosing initial conditions. It is moreover more
deeply connected to modifications of gravity (rather than just clustering
of dark energy) than $Y$ \cite{Saltas:2014dha}. Due to these properties,
$\eta$ can be estimated by future clustering and lensing data (or
even from B-modes of the cosmic microwave background data \cite{Amendola:2014wma,Raveri:2014eea})
to a precision of just one percent if assumed constant \cite{Amendola:2013qna}.
This is to be contrasted to the 30\% relative errors that can be obtained
on the combination $\Omega_{m,0}Y$ when model-independency is at
least partially taken into account.

\section{The Horndeski parameters}

We are interested in the evolution of linear perturbations in the
quasi-static limit (i.e. for scales significantly inside the cosmological
horizon, $k/(aH)\gg1$, and inside the Jeans lenght of the scalar,
$c_{s}k/(aH)\gg1$, such that the terms containing $k$ dominate over
the time-derivative terms). One has then the following equation of
linear perturbation growth 
\begin{equation}
\delta''_{m}+(2+\frac{E'}{E})\delta'_{m}=\frac{3}{2}\Omega_{m}\delta_{m}Y(a,k)\label{delta}
\end{equation}
where $\Omega_{m}=\Omega_{m0}a^{-3}/E^{2}$ and $E\equiv H/H_{0}$.
The prime denotes the derivative respect to $N\equiv\ln(a)$. The
function $Y$, the effective gravitational constant for matter, is
defined as 
\begin{equation}
Y(a,k)=-\frac{2k^{2}\Psi}{3(aH)^{2}\Omega_{m}\delta_{m}}\label{Y}
\end{equation}
In this paper we always assume either that baryons do not feel modifed
gravity or that the local gravity experiments occur in an environment
where the extra force is not felt; in either case, they do not set
any useful constraint on the cosmological expression for $Y$.

Now we assume that the background is described by the $\Lambda$CDM
model, so we have: 
\begin{equation}
E^{2}=\Omega_{m,0}^{(bg)}a^{-3}+1-\Omega_{m,0}^{(bg)}\label{back}
\end{equation}
where we distinguish here between the parameter $\Omega_{m,0}^{(bg)}$
that enters the background rate $E$ and the parameter $\Omega_{m,0}$
that expresses the amount of clustered matter in Eq. (\ref{delta}).
In a modified gravity theory, the two quantities are independent and
should be clearly distinguished. A perfect knowledge of the expansion
rate, e.g. through supernovae Ia, will determine $E$ and, if the
particular form (\ref{back}) is assumed, $\Omega_{m,0}^{(bg)}$ but
says nothing about the clustered fraction of matter $\Omega_{m,0}$.
For instance, if dark energy mediates an extra force, matter will
not dilute as $a^{-3}$ and the value of $\Omega_{m,0}^{(bg)}$ that
one would obtain from (\ref{back}) would be unrelated to the real
matter content.

The Horndeski Lagrangian is the most general Lagrangian for a single
scalar field which gives second-order equations of motion for both
the scalar field and the metric on an arbitrary background. In the
quasi-static limit of the Horndeski Lagrangian one obtains: 
\begin{equation}
Y=h_{1}\frac{1+(k/k_{p})^{2}h_{5}}{1+(k/k_{p})^{2}h_{3}}\label{Y1}
\end{equation}
where $h_{1},h_{3},h_{5}$ are time dependent functions that can be
explicitly obtained when the full Horndeski Lagrangian is given \cite{2013PhRvD..87b3501A}.
The scale $k_{p}$ is an arbitrary pivot scale that we choose to be
$k_{p}=1h$/Mpc.

Eq. (\ref{delta}) can be written as

\begin{equation}
\delta''_{m}+(2+\frac{E'}{E})\delta'_{m}=\frac{3}{2}\frac{\delta_{m}}{a^{3}E^{2}}\Omega_{m,0}Y\label{eq:a-1-1}
\end{equation}
This shows immediately that $\Omega_{m,0}$ is fully degenerate with
$Y$. In the following therefore we will only be able to constrain
the quantity 
\begin{equation}
\hat{Y}\equiv\Omega_{m,0}Y
\end{equation}
Since our reference model is $\Lambda$CDM with $\Omega_{m,0}^{(bg)}=0.3$,
the standard value of $\hat{Y}$ is 0.3. Similarly, when we take the
specific Horndeski form (\ref{Y1}), we will constrain the combination
$\hat{h}_{1}\equiv\Omega_{m,0}h_{1}$.

Now, the rate $E$ itself can be estimated with distance indicators
only up to some uncertainty. In the following however we will simplify
our task by assuming that the error on $E$ is actually already now
negligible with respect to the errors on the other observational data.
For current data this is not completely true so our estimate of the
uncertainty on $Y$ is actually a lower limit. As the main effect
of a change in $E$ is through the left-hand-side factor $E^{-2}$
in Eq. (\ref{eq:a-1-1}), one can estimate the additional error on
$\hat{Y}$ induced by an error $\Delta E$ in $E$ to be $|\Delta\hat{Y}|/\hat{Y}\approx2|\Delta E|/E$,
to be added in quadrature. Current supernovae can determine $E$ around
$z\approx1$ with a relative error of 5-10\%, so we can estimate an
additional error on $\hat{Y}$ around 10-20\%. Since the uncertainty
we find is quite larger than this, we neglect the additional source
of error from $E$. For the future data, one can indeed assume that
$E$ will be pretty fairly well determined up to better than a percent
accuracy with future surveys  and our lower limit will be closer to
reality.

For the initial conditions on $\delta_{m}$, we fix the irrelevant
value $\delta_{in}=e^{N_{in}}$ with $N_{in}=-1.5$ i.e. $z_{in}\approx3.5$,
while for the initial growth rate parameter $\alpha=\delta'_{in}/\delta_{in}$
we either fix it to unity (standard $\Lambda$CDM) or adopt a uniform
prior large enough to cover all the region in which the likelihood
is significantly different from zero.

Two caveats are in order. First, the entire analysis of this paper
deals with linear scales. However, the data points we employ are obtained
averaging over various scales that include probably also some weakly
non-linear region of the power spectrum. For instance, the effective
wavenumber in the analysis of Ref. \cite{Beutler:2013yhm} is given
as $k_{\mathrm{eff}}=0.178h/$Mpc, which at the average redshift of
0.57 is marginally affected by non linearity. In Ref. \cite{Samushia:2013yga}
the analysis of the same data including only large linear scales leads
to an estimate of $f\sigma_{8}(z)$ which is consistent with, and
only mildly more uncertain than, the one obtained including smaller
scales, indicating that the non linear effects are still subdominant.
In any case, properly dealing with non linearity would require a reanalysis
of the raw clustering data and an estimate of the non-linear corrections
to $\delta_{m}$ for non-standard model. Secondly, the data points
have been obtained by assuming a particular background expansion in
order to convert from redshift to distances. Here we assume a fiducial
$\Lambda$CDM background with $\Omega_{m,0}=0.3$ which does not coincide
exactly with the one employed in some of the real data analysis. The
corrections induced by both the non-linear effects and the fiducial
background mismatch are expected to be quite smaller than the rather
large error that we obtain on our modified gravity parameters.

\section{Marginalization over $\sigma_{8}$}

We build a data posterior by using two datasets, the \emph{current
dataset} and the \emph{forecast dataset}. The current dataset includes
all the independent published estimates of $f\sigma_{8}(z)$ obtained
with the redshift distortion method. It includes the data from 2dFGS,
6dFGS, LRG, BOSS, CMASS, WiggleZ and VIPERS, and spans the redshift
interval from $z=0.07$ to $z=0.8$, see Table \ref{tab:Current-published-values}
(see also \cite{Macaulay:2013swa,More:2014uva}). In some case the
correlation coefficient between two samples has been estimated in
Ref. \cite{Macaulay:2013swa} and included in our analysis; when there
are different published results from the same dataset in Table \ref{tab:Current-published-values}
we include only the more recent one. The forecast dataset approximates
instead the accuracy of a future Euclid mission \cite{Euclid-r,Amendola:2012ys}
and it has been obtained in Ref. \cite{Amendola:2013qna} in the range
from $z=0.5$ to $z=2.1$. The growth rate data are given as a set
of values $d_{i}$ at various redshifts, where 
\begin{equation}
d_{i}=f(z_{i})\sigma_{8}(z_{i})=f(z_{i})\sigma_{8}G(z_{i})=\sigma_{8}\frac{\delta'}{\delta_{0}}\label{d-1}
\end{equation}
and where $f(z)=\delta'_{m}/\delta_{m}$ is the growth rate, $G(z)$
is the growth factor normalized to unity today and $\sigma_{8}$ is
the present power spectrum normalization. We denote our theoretical
estimates as $t_{i}=\delta'_{i}/\delta_{0}$. We build then the $\chi^{2}$
function 
\begin{equation}
\bar{\chi}_{f\sigma_{8}}^{2}=(d_{i}-\sigma_{8}t_{i})C_{ij}^{-1}(d_{j}-\sigma_{8}t_{j})\label{chiquadro-1}
\end{equation}
where $C_{ij}$ is the covariant matrix of the data. The first step
to implement our model-independent estimates is to marginalize over
$\sigma_{8}$, since as already mentioned to estimate its value from
current data one would need to know the gravitational theory. Marginalizing
the likelihood $L'=\exp(-\bar{\chi}_{f\sigma_{8}}^{2}/2)$ over $\sigma_{8}>0$
with uniform prior leads to a new posterior $L=\exp(-\bar{\chi}_{f\sigma_{8}}^{2}/2)$
where 
\begin{equation}
\chi_{f\sigma_{8}}^{2}=S_{dd}-\frac{S_{dt}^{2}}{S_{tt}}+\log S_{tt}-2\log(1+\mathrm{Erf}(\frac{S_{dt}}{\sqrt{2S_{tt}}}))\label{chimarg-1}
\end{equation}
and where 
\begin{eqnarray}
S_{dt} & = & d_{i}C_{ij}^{-1}t_{j}\\
S_{dd} & = & d_{i}C_{ij}^{-1}d_{j}\\
S_{tt} & = & t_{i}C_{ij}^{-1}t_{j}
\end{eqnarray}
This is the posterior distribution we will use in the following discussion.

\begin{center}
\begin{table}
\centering{}\protect%
\begin{tabular}{|c|c|c|c|}
\hline 
Survey  & z  & $f(z)$$\sigma_{8}(z)$  & References\tabularnewline
\hline 
\hline 
6dFGRS  & 0.067  & 0.423 \textpm{} 0.055  & Beutler et al. (2012) \cite{Beutler:2012px}\tabularnewline
\hline 
\multirow{2}{*}{LRG-200} & 0.25  & 0.3512 \textpm{} 0.0583  & \multirow{1}{*}{Samushia et al (2012) \cite{Samushia:2011cs}}\tabularnewline
\cline{2-3} 
 & 0.37  & 0.4602 \textpm{} 0.0378  & \tabularnewline
\hline 
\multirow{2}{*}{LRG-60} & 0.25{*}  & 0.3665$\pm$0.0601  & Samushia et al (2012) \cite{Samushia:2011cs}\tabularnewline
\cline{2-3} 
 & 0.37{*}  & 0.4031$\pm$0.0586  & \tabularnewline
\hline 
\multirow{2}{*}{BOSS} & 1) 0.30  & 0.408$\pm$ 0.0552, $\rho_{12}=-0.19$  & Tojeiro et al. (2012)\cite{Tojeiro:2012rp}\tabularnewline
\cline{2-3} 
 & 2) 0.60  & 0.433$\pm$ 0.0662  & \tabularnewline
\hline 
\multirow{3}{*}{WiggleZ} & 1) 0.44  & 0.413 \textpm{} 0.080, $\rho_{12}=0.51$  & \multirow{1}{*}{Blake (2011) \cite{Blake:2012pj}}\tabularnewline
\cline{2-3} 
 & 2) 0.60  & 0.390 \textpm{} 0.063, $\rho_{23}=0.56$  & \tabularnewline
\cline{2-3} 
 & 3) 0.73  & 0.437 \textpm{} 0.072  & \tabularnewline
\hline 
Vipers  & 0.8  & 0.47 \textpm{} 0.08  & De la Torre et al (2013)\cite{delaTorre:2013rpa}\tabularnewline
\hline 
2dFGRS  & 0.13  & 0.46 \textpm{} 0.06  & Percival et al. (2004) \cite{Percival:2004fs}\tabularnewline
\hline 
LRG  & 0.35  & 0.445 \textpm{} 0.097  & Chuang and Wang (2013) \cite{Chuang:2012qt}\tabularnewline
\hline 
LOWZ  & 0.32  & 0.384$\pm$0.095  & \multirow{1}{*}{Chuang at al (2013)\cite{Chuang:2013wga}}\tabularnewline
\cline{1-3} 
\multirow{4}{*}{CMASS} & 0.57{*}  & 0.348 \textpm{} 0.071  & \multirow{1}{*}{}\tabularnewline
\cline{2-4} 
 & 0.57{*}  & 0.423 \textpm{} 0.052  & Beutler et al (2014)\cite{Beutler:2013yhm}\tabularnewline
\cline{2-4} 
 & 0.57  & 0.441$\pm$0.043  & Samushia et al (2014) \cite{Samushia:2013yga}\tabularnewline
\cline{2-4} 
 & 0.57{*}  & 0.450 \textpm{} 0.011  & Reid et al (2013)\cite{Reid:2014iaa}\tabularnewline
\hline 
\end{tabular}\caption{\label{tab:Current-published-values}Current published values of $f\sigma_{8}(z)$.
In some cases we list also the correlation coefficient $\rho_{ij}$
between different bins \cite{Macaulay:2013swa}. Entries with an asterisk
are not employed in this analysis. }
\end{table}

\par\end{center}

\section{Current growth-rate data}

Current growth data are not sufficient to provide $k$-dependent information.
In this case, therefore, we are forced to neglect the $k$-dependence
of $Y$. Moreover, again in view of the lack of sufficient statistics,
we also fix the time dependence and assume that $Y$ is just a constant
over the redshift range of the observations. We have therefore just
two parameters: $\hat{Y}$ and the initial condition $\alpha$. We
will consider four cases, in increasing order of ``model independence''.
The \emph{first case} is standard $\Lambda$CDM gravity ($Y=1$),
and $\sigma_{8}$ and initial conditions both fixed to the fiducial
model, $\sigma_{8}=0.83$ \cite{Planck_016} and $\alpha=1$. Here
the only free parameter is therefore $\Omega_{m,0}$ (with an uniform
prior in $0,1$). The \emph{second case} is like the first one but
with marginalization over $\sigma_{8}$. This case serves mainly to
isolate the effect of the $\sigma_{8}$-marginalization and to see
how much the best fit of $\Omega_{m,0}$ changes if $\sigma_{8}$
is estimated from the $f\sigma_{8}(z)$ data themselves and not from
Planck. From now on, we always fix the background evolution $E$ to
a $\Lambda$CDM with $\Omega_{m,0}^{(bg)}$ = 0.3, in agreement with
observations and close to the Planck best fit \cite{Planck_016},
and neglecting any uncertainty on it and we always include the marginalization
over $\sigma_{8}$. In the \emph{third case}, beside marginalizing
over $\sigma_{8}$, we leave $\hat{Y}$ free to vary with an uniform
prior for positive values. Finally, the \emph{fourth case} is like
the third one but now the initial growth rate $\alpha$ is left free
to vary.

\begin{figure}[t]
\includegraphics[width=7cm]{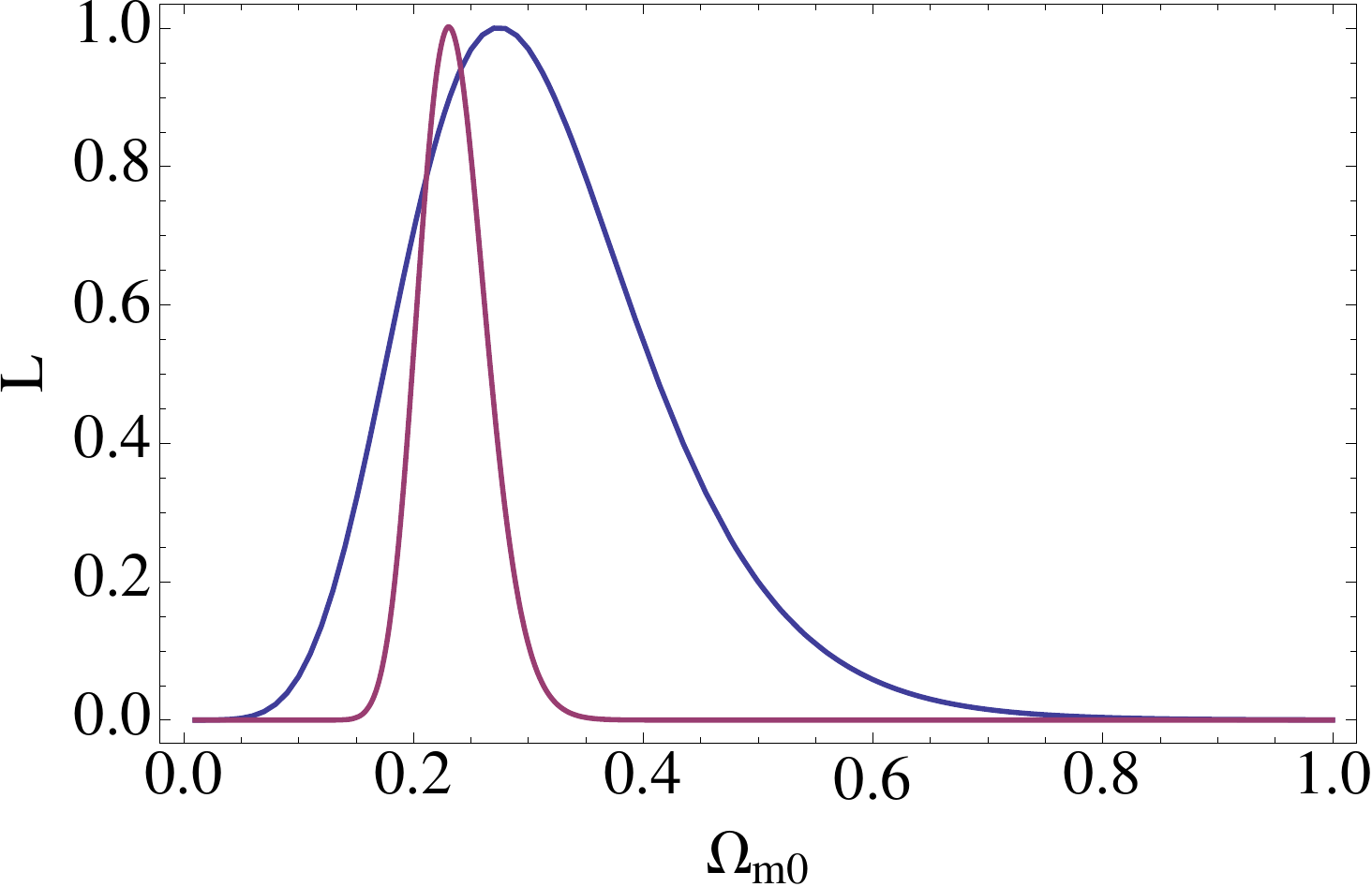} \includegraphics[width=7cm]{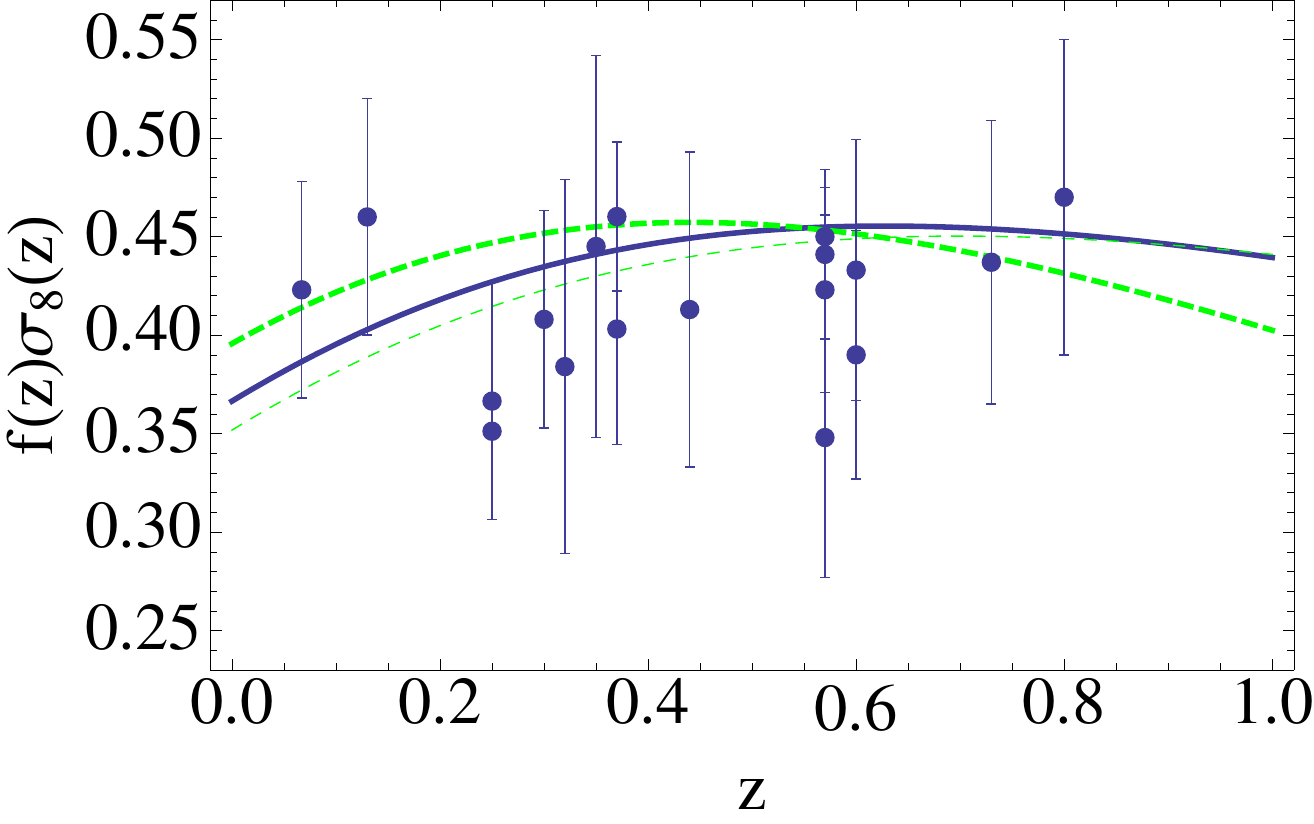}

\caption{$\emph{Left panel}:$\emph{ }Current posterior in function of $\Omega_{m,0}$
for $\Lambda$CDM model (first case). $\emph{Right panel}:$ Best
fit $\Lambda$CDM model for the first case (blue solid curve); best
fit for the third case (green thin dashed curve) and for the fourth
case (green thick dashed curve) together with the entire set of $f\sigma_{8}$
data points we employed in this paper. As the posterior is marginalized
over $\sigma_{8}$, a possible vertical rescaling for the third and
fourth cases is inconsequential so they have been plotted with a normalization
that minimizes the $\chi^{2}$ distance.}

\label{Fig1} 
\end{figure}

The results for the first and second cases are shown on the left panel
of Fig. (\ref{Fig1}). At 68\% c.l., the uncertainty on $\Omega_{m,0}$
increases from 0.03 to roughly 0.10 when marginalizing over $\sigma_{8}$
while for $\sigma_{8}$ itself we find $\sigma_{8}=0.76_{-0.10}^{+0.06}$,
smaller than but compatible with the Planck value. On the right panel,
we plot the $f\sigma_{8}$ data points from the galaxy surveys with
their respective error bars in comparison with the $\Lambda$CDM model
and with the best fits of all the cases. 

\begin{figure}[t]
\includegraphics[bb=0bp 0bp 240bp 239bp,width=6.6cm]{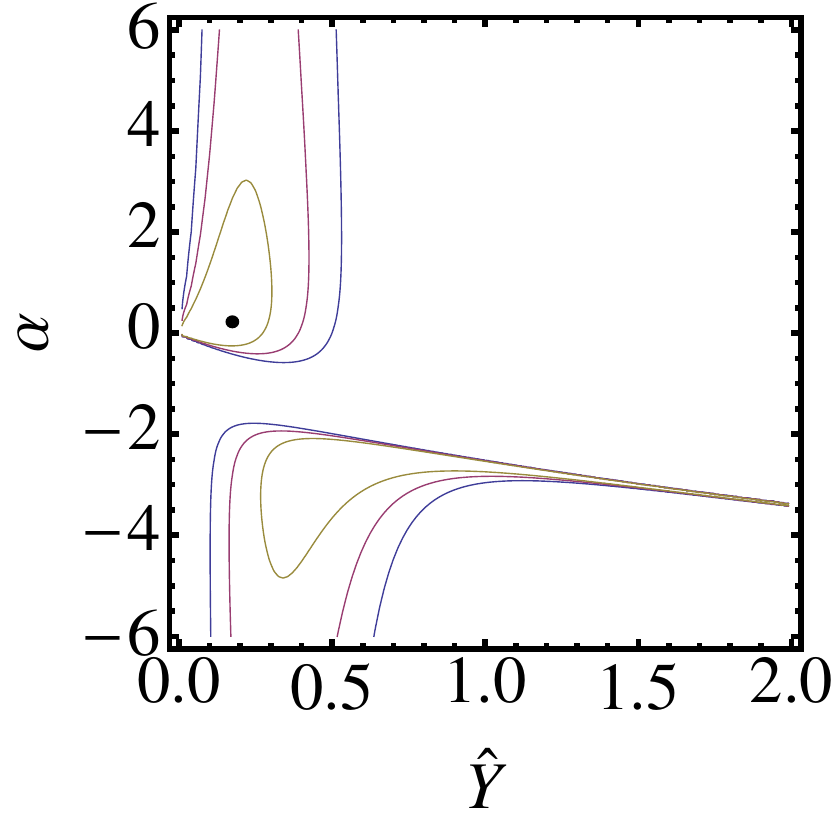}

\includegraphics[width=6.6cm]{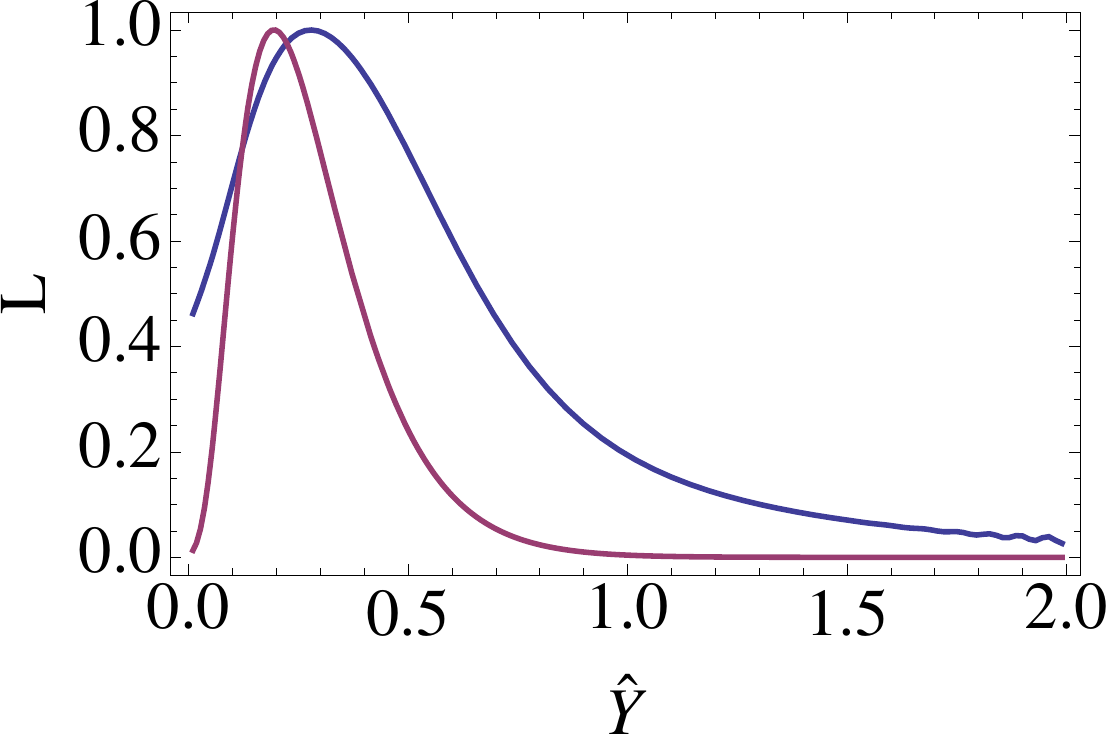} 
\includegraphics[bb=0bp 0bp 240bp 168bp,width=6.6cm]{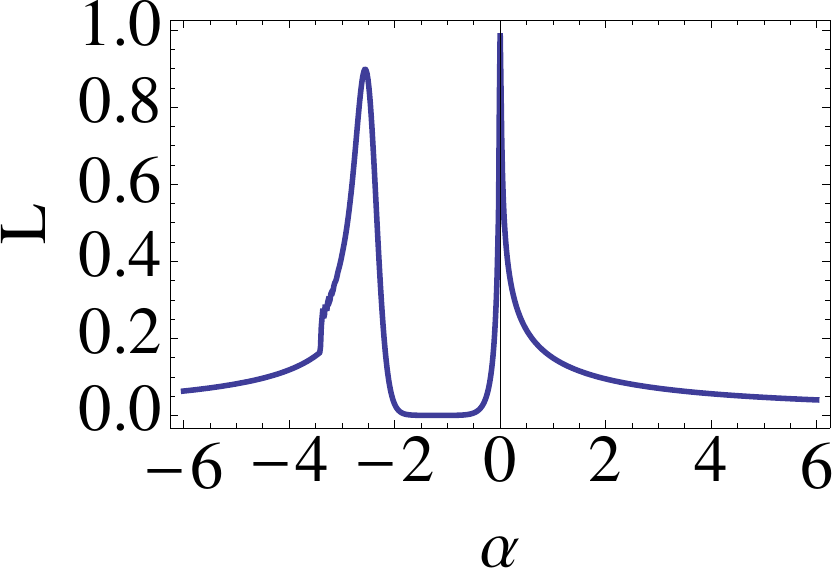}

\caption{$\emph{Top panel }$: 1$\sigma$, 2$\sigma$ and 3$\sigma$ confidence-level
contours for the 2-dimensional current posterior on the parameters
$\{\hat{Y},$$\alpha\}$ marginalizing over $\sigma_{8}$ (fourth
case). $\emph{Left bottom panel }$:\emph{ }Current posterior for
$\hat{Y}$ marginalized over $\alpha$ (blue line) in comparison with
the third case (red line). $\emph{Right bottom panel }$: Current
posterior for $\alpha$ marginalized over $\hat{Y}$.}

\label{Fig2} 
\end{figure}

\begin{figure}[t]
\includegraphics[bb=0bp 0bp 492bp 489bp,width=0.5\textwidth]{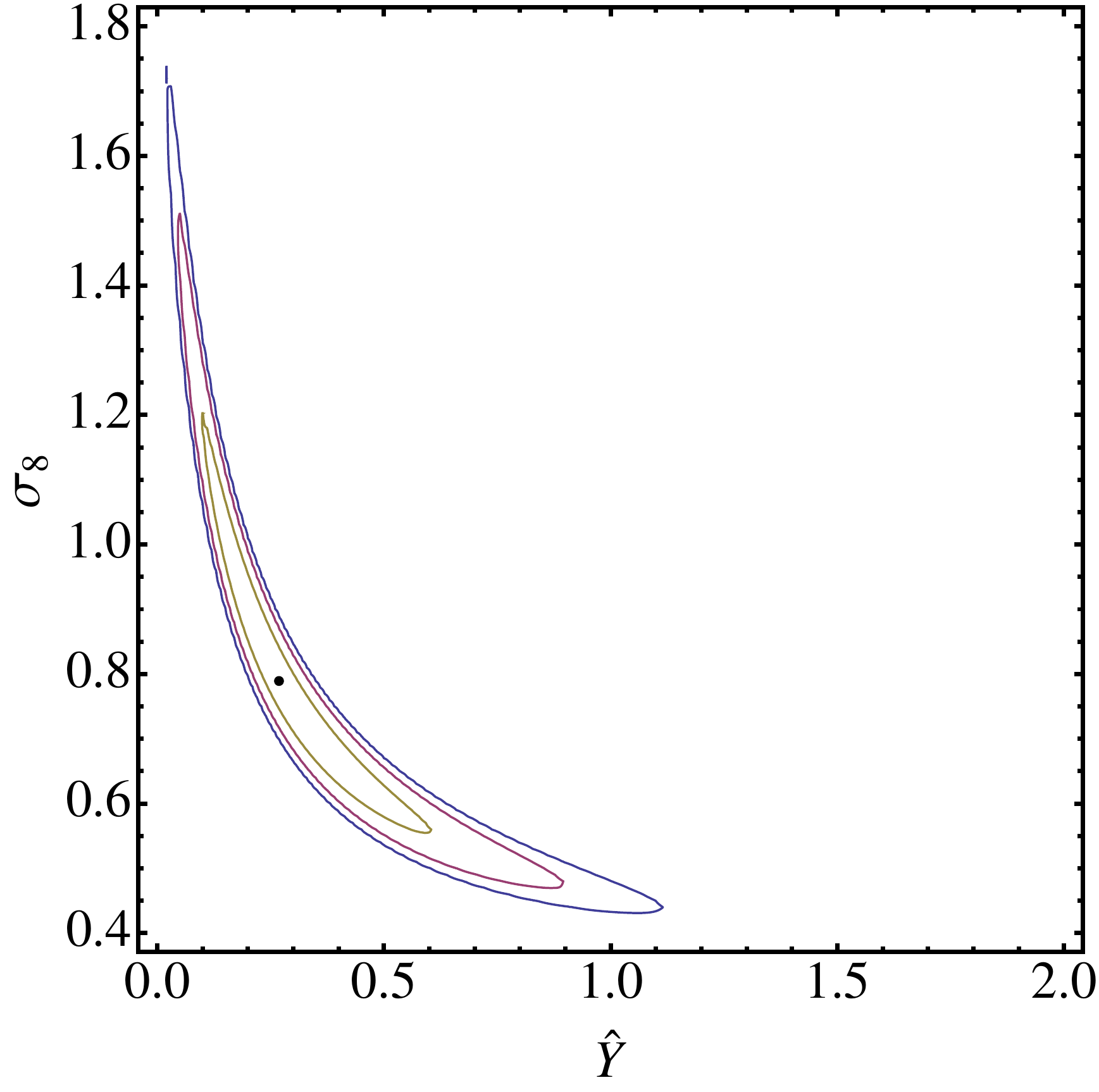}

\caption{1$\sigma$, 2$\sigma$ and 3$\sigma$ confidence-level contours for
the current posterior on the parameters $\{\hat{Y},$$\sigma_{8}\}$
(third case).}

\label{fig:case3-c-y-s8} 
\end{figure}

In the third case (uniform prior on $\hat{Y}$, fixing $\alpha=1$)
we find a best fit $\hat{Y}=0.20$ with an error range $[0.095,0.36]$
at 68\% confidence level, see Fig. (\ref{Fig2}) bottom left panel.
The parameter $\sigma_{8}$ is now $\sigma_{8}=0.79_{-0.25}^{+0.11}$,
see Fig. (\ref{fig:case3-c-y-s8}). If now we vary $\alpha$ with
a uniform prior on $\hat{Y}$ (fourth case, Fig. \ref{Fig2}) we obtain
instead $\hat{Y}=0.28$ with a doubled error range $[0.048,0.63]$
at 68\% c.l. . For $\sigma_8$  we have now 
$\sigma_{8}=0.54_{-0.09}^{+0.21}$.
Table (\ref{tab:Summary-of-results-1}) summarizes the
results.

Interestingly, we detect a bimodality in the marginalized posterior
for $\alpha$ and a strong correlation with $\hat{Y}$. As shown in
Fig. (\ref{Fig2}) both very large and very small values of $\hat{Y}$
are acceptable if $\alpha$ varies freely. In particular, a large
$\hat{Y}$ can be compensated by a large negative $\alpha$, while
small $\hat{Y}$ are compatible with large positive values. Large
negative values of $\alpha$ mean that overdensities can become underdensities
at some point in time; although this might appear pathological at
first sight, it does not contradict any observation at linear scales
and should not be arbitrarily excluded. Within 3$\sigma$, a small
$\hat{Y}$ is compatible with any value of $\alpha$ since in this
limit the perturbation equation becomes effectively first order in
$\delta'$.

The conclusion of this section is that current data put hardly any
constraint on $\hat{Y}$. Any value from 0 to 1.35 is acceptable at
95\% and much larger values of $\hat{Y}$ are also acceptable if the
initial condition is chosen along the degeneracy line of Fig. (\ref{Fig2}).

This conclusion could have been reasonably expected due to the
paucity of present data. In the next section we show however that
 the constraints
improve a lot  with the much better data of
future surveys only if we keep $\alpha$ fixed; in the more general case,
the improvement remains modest.
The reason is the same: trying to be as much model-independent
as possible one has to set $\sigma_{8},\Omega_{m,0},\alpha$ free
to vary. The price to pay for this freedom are rather weak constraints.

\begin{table}
\begin{tabular}{|c|c|c|c|c|c|c|c|}
\hline 
\multirow{2}{*}{} & \multirow{2}{*}{case } &  $\alpha$  & \multirow{2}{*}{$\Delta\alpha$ (95\%) } & \multirow{2}{*}{$\Delta\alpha$ (68\%)} &  $\Omega_{m,0}^{(bg)}$  & \multirow{2}{*}{$\Delta\Omega_{m,0}^{(bg)}$ ($95\%$) } & \multirow{2}{*}{$\Delta\Omega_{m,0}^{(bg)}$ ($68\%$ )}\tabularnewline
 &  & (best fit) &  &  & (best fit) &  & \tabularnewline
\hline 
\multirow{2}{*}{$\Lambda$CDM, $Y=1$, $\sigma_{8}=0.83$ } & \multirow{2}{*}{I } & \multirow{2}{*}{1 } & \multirow{2}{*}{-} & \multirow{2}{*}{-} & \multirow{2}{*}{ 0.23} & \multirow{2}{*}{ {[}0.18, 0.29{]}} & \multirow{2}{*}{{[}0.20, 0.26{]}}\tabularnewline
 &  &  &  &  &  &  & \tabularnewline
\hline 
\multirow{2}{*}{$\Lambda$CDM, $Y=1$, marg. on $\sigma_{8}$ } & \multirow{2}{*}{II} & \multirow{2}{*}{1} & \multirow{2}{*}{-} & \multirow{2}{*}{-} & \multirow{2}{*}{ 0.27} & \multirow{2}{*}{{[}0.12, 0.54{]}} & \multirow{2}{*}{{[}0.18, 0.39{]}}\tabularnewline
 &  &  &  &  &  &  & \tabularnewline
\hline 
\multirow{2}{*}{} & \multirow{2}{*}{case} & \multirow{2}{*}{} & \multirow{2}{*}{} & \multirow{2}{*}{} & \multirow{2}{*}{$\hat{Y}$ } & \multirow{2}{*}{$\Delta\hat{Y}$ ($95\%$) } & \multirow{2}{*}{$\Delta\hat{Y}$ ($68\%$)}\tabularnewline
 &  &  &  &  &  &  & \tabularnewline
\hline 
\multirow{2}{*}{Uniform prior on $\hat{Y}$ } & \multirow{2}{*}{III} & \multirow{2}{*}{1} & \multirow{2}{*}{-} & \multirow{2}{*}{-} & \multirow{2}{*}{0.20} & \multirow{2}{*}{ {[}0.040, 0.60{]}} & \multirow{2}{*}{{[}0.095, 0.36{]}}\tabularnewline
 &  &  &  &  &  &  & \tabularnewline
\hline 
\multirow{2}{*}{Uniform prior on $\hat{Y},\alpha$ } & \multirow{2}{*}{IV } & \multirow{2}{*}{-0.015 } & $\leq$-2.08 and & {[}-0.40, 1.32{]} and  & \multirow{2}{*}{0.28} & \multirow{2}{*}{{[}0, 1.35{]} } & \multirow{2}{*}{{[}0.048, 0.63{]}}\tabularnewline
 &  &  & $\geq$ -0.67 & {[}-4.05, -2.20{]} &  &  & \tabularnewline
\hline 
\end{tabular}

\caption{Summary of results for current data.}

\label{tab:Summary-of-results-1} 
\end{table}

\begin{figure}
\includegraphics[width=6.6cm]{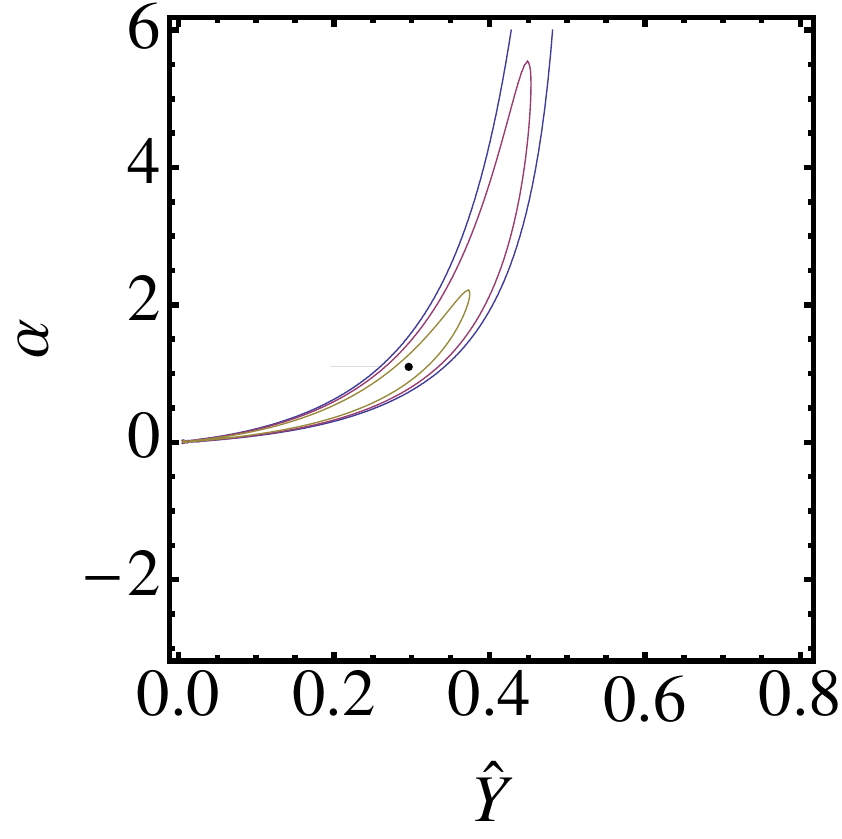}

\includegraphics[width=6.6cm]{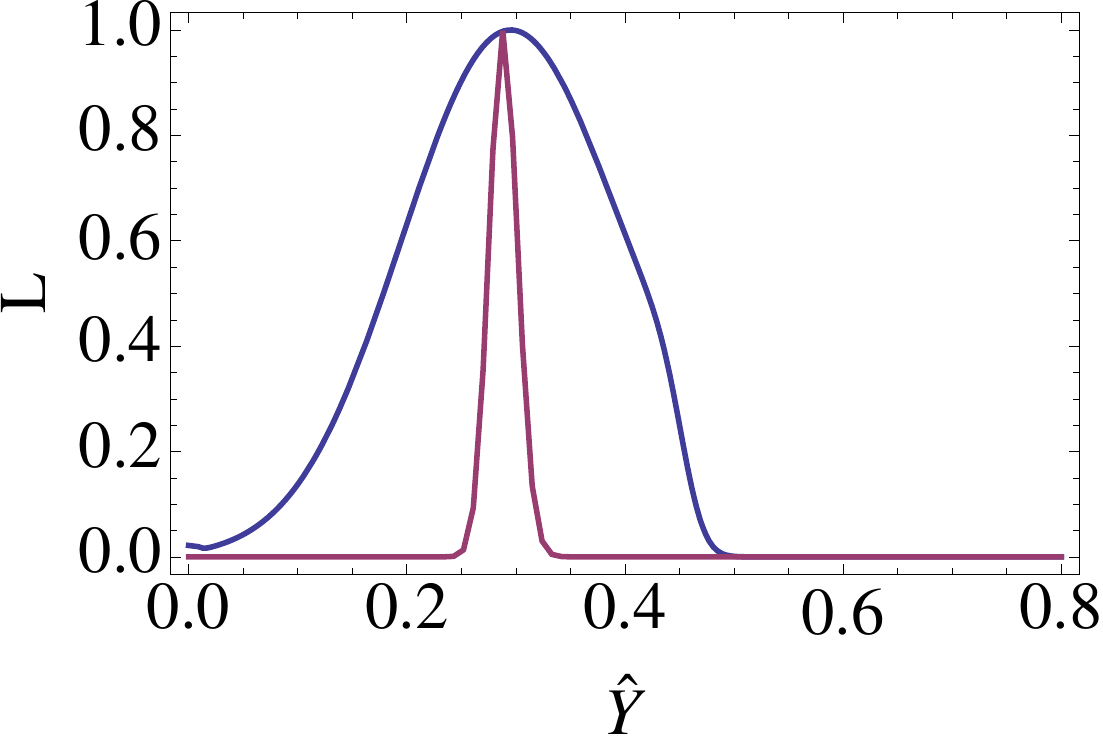}\includegraphics[width=6.6cm]{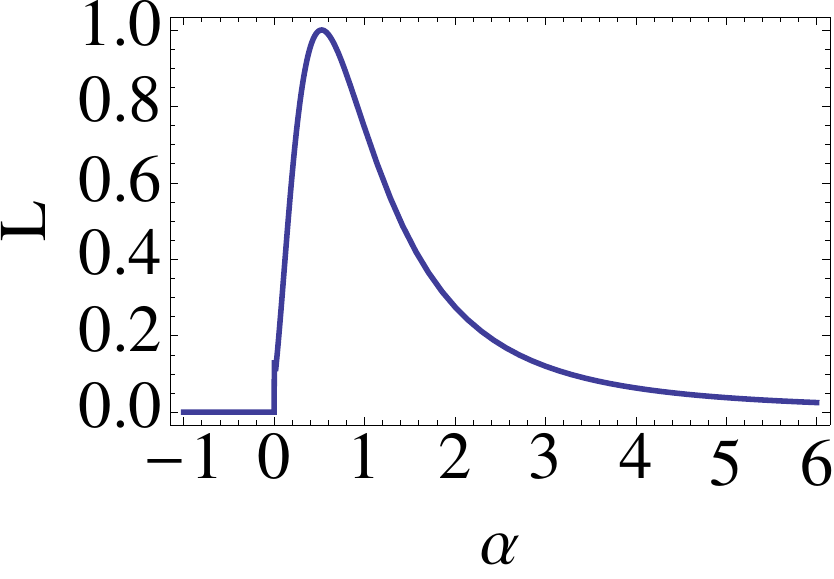}

\caption{\label{fig:case4-f}$\emph{Top panel }$: 1$\sigma$, 2$\sigma$ and
3$\sigma$ confidence-level contours for the 2-dimensional forecast
posterior on the parameters $\{\hat{Y},$$\alpha\}$ marginalizing
over $\sigma_{8}$ (fourth case). $Left\emph{ bottom panel }$: forecast
posterior for $\hat{Y}$ marginalized over $\alpha$ (blue line) in
comparison with the third case (red line). $\emph{Right bottom panel }$:
forecast posterior for $\alpha$ marginalized over $\hat{Y}$}
\end{figure}

\begin{figure}
\includegraphics[width=0.5\textwidth]{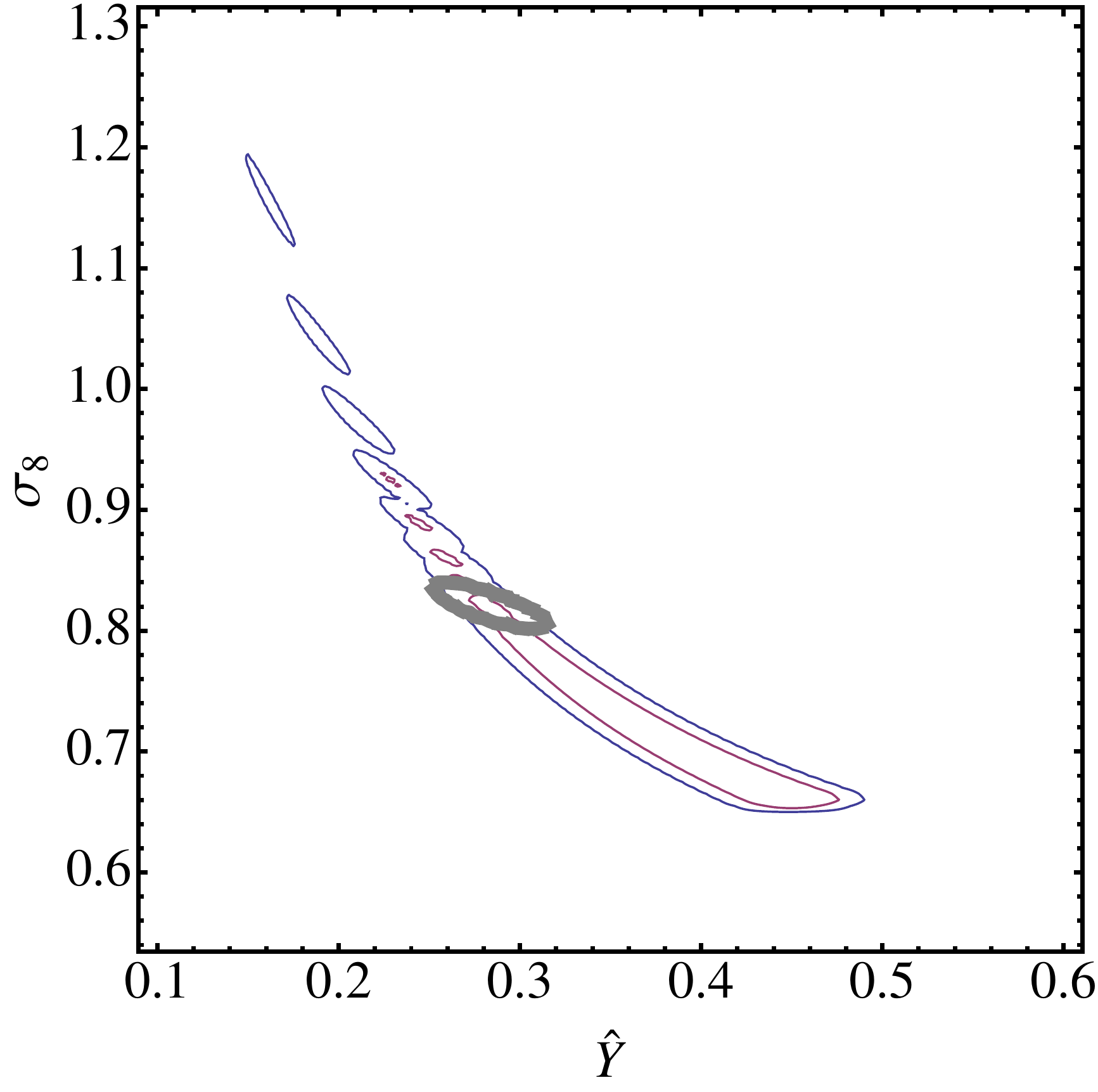}

\caption{\label{fig:case4-f-y-s8}1$\sigma$ and 2$\sigma$ confidence-level
contours for the forecast posterior on the parameters $\{\hat{Y},$$\sigma_{8}\}$
when $\alpha$ is marginalized over (fourth case, solid curves) and
when is fixed (third case, small gray ellipse, only the  2$\sigma$ contour is shown).}

\end{figure}

\section{Forecast data}

\section*{A. $z$ binning }

In this section we consider the forecast Euclid-like $f\sigma_{8}$
datasets, starting with the case of no scale information ($z$-binning),
which can be directly compared to the previous ones. The growth forecasts
are obtained from Ref. \cite{Amendola:2013qna}. We consider a Euclid-like
15,000 square degrees redshift survey from $z=0.5-1.5$ divided in
equally spaced bins of width $\Delta z=0.2$ and, in order to prevent
accidental degeneracy due to low statistic, a single larger redshift
bin between $z=1.5-2.1$, so in total we have six bins. In Table \ref{tab:table1}
we show the fiducial values and relative errors on $f\sigma_{8}$
.

\begin{table}[htbp]
\begin{centering}
\begin{tabular}{|l|c|c|}
\hline 
$\bar{z}$  & $f\sigma_{8}$  & $\Delta f\sigma_{8}$ (68\% c.l.)\tabularnewline
\hline 
0.6  & 0.469  & 0.0092\tabularnewline
\hline 
0.8  & 0.457  & 0.0068\tabularnewline
\hline 
1.0  & 0.438  & 0.0056\tabularnewline
\hline 
1.2  & 0.417  & 0.0049\tabularnewline
\hline 
1.4  & 0.396  & 0.0047\tabularnewline
\hline 
1.8  & 0.354  & 0.0039\tabularnewline
\hline 
\end{tabular}
\par\end{centering}

\protect\protect\protect\protect\caption{Fiducial values and Euclid-like errors for $f\sigma_{8}$ using six
redshift bins (from \cite{Amendola:2013qna}). }

\label{tab:table1} 
\end{table}

As before, we want to obtain an estimate on a constant $\hat{Y}$
marginalizing over $\sigma_{8}$ and $\alpha$. Fig. (\ref{fig:case4-f}),
lower left panel, shows the 1-dimensional marginalized \emph{forecast
posterior} distribution of $Y$ (third case) along with the fourth
case, i.e. with marginalization over $\alpha$. As can be seen from
Fig. (\ref{fig:case4-f}), lower left panel, the 95\% error on $\hat{Y}$
around the fiducial value 0.3 has a fivefold increase, from $0.03$
to roughly $0.15$, when we marginalize over the initial conditions.
The relative uncertainly on $\hat{Y}$ is around 30\% at 68\% c.l..
Contrary to what we found previously using current data, negative
values of $\alpha$ appear now strongly disfavoured.

The increase in errors on both $\hat{Y}$ and $\sigma_{8}$ can be
appreciated from Fig. (\ref{fig:case4-f-y-s8}). In the third case
(i.e. no marginalization over $\alpha$) future Euclid-like data can
estimate $\sigma_{8}$ and $\hat{Y}$ to within 0.01 for both parameters;
when $\alpha$ is marginalized over however the error increases to
roughly $0.08$, again for both parameters.

\begin{table}
\protect%
\begin{tabular}{|c|c|c|c|c|c|c|c|}
\hline 
 & case  & Best fit $\alpha$  & $\Delta\alpha$ (95\%)  & $\Delta\alpha$ (68\%)  & Best fit $\hat{Y}$  & $\Delta Y$ ($95\%$)  & $\Delta Y$ ($68\%$)\tabularnewline
\hline 
\hline 
Uniform prior on $\hat{Y}$  & III  & 1  & - & - & 0.29  & {[}0.26, 0.32{]}  & {[}0.28, 0.30{]} \tabularnewline
\hline 
Uniform prior on $\hat{Y},\alpha$  & IV  & 0.53  & {[}0, 4.0{]}  & {[}0.12, 1.6{]}  & 0.30  & {[}0.12, 0.43{]}  & {[}0.21, 0.38{]}\tabularnewline
\hline 
\end{tabular}\protect\caption{Summary of results for forecasted Euclid data.}

\label{tab3} 
\end{table}

\section*{B. $k$ binning}

We consider now the quasi-static Horndeski result, defined in Eq.
(\ref{Y1}), which contains the parameters $h_{1}$, $h_{3}$ and
$h_{5}$ and a $k-$dependence. Although in general these parameters
depend on time, we assume here for simplicity that they time variation
is negligible in the observed range. The aim of this section is to
obtain error estimates on the Horndeski parameters, so we need to
have a minimum of three $k$-bins for every value of the redshift.
Again following the method of \cite{Amendola:2013qna} we take the
minimum binning value of $k$ as $k_{min}=0.007$ $h/Mpc$ (the result
is very weakly dependent on this value) and the values of the highest
$k$ are chosen to be well below the scale of non-linearity at the
redshift of the bin. In Table \ref{tab4} we report the $k$-bin boundaries.

In Table \ref{tab7} we display the fiducial values and errors for
$f\sigma_{8}$ at every redshift and every $k$- bin. As in the previous
case, also here the fiducial model is chosen to be $\Lambda$CDM,
so the fiducial values for the Horndeski parameters are $\hat{h}_{1}=\Omega_{m,0}h_{1}=0.3$
and $h_{3}=h_{5}=0$. Here we fix $h_{5}$ to its fiducial value (i.e.
to zero) due to the degeneracy between $h_{5}$ and $h_{3}$ when
the fiducial model is such that $h_{5}$ = $h_{3}$ as in $\Lambda$CDM.
In the next section we will consider the case in which the fiducial
value of $h_{5}$ is different from the standard value.

\begin{figure}
\includegraphics[width=6.6cm]{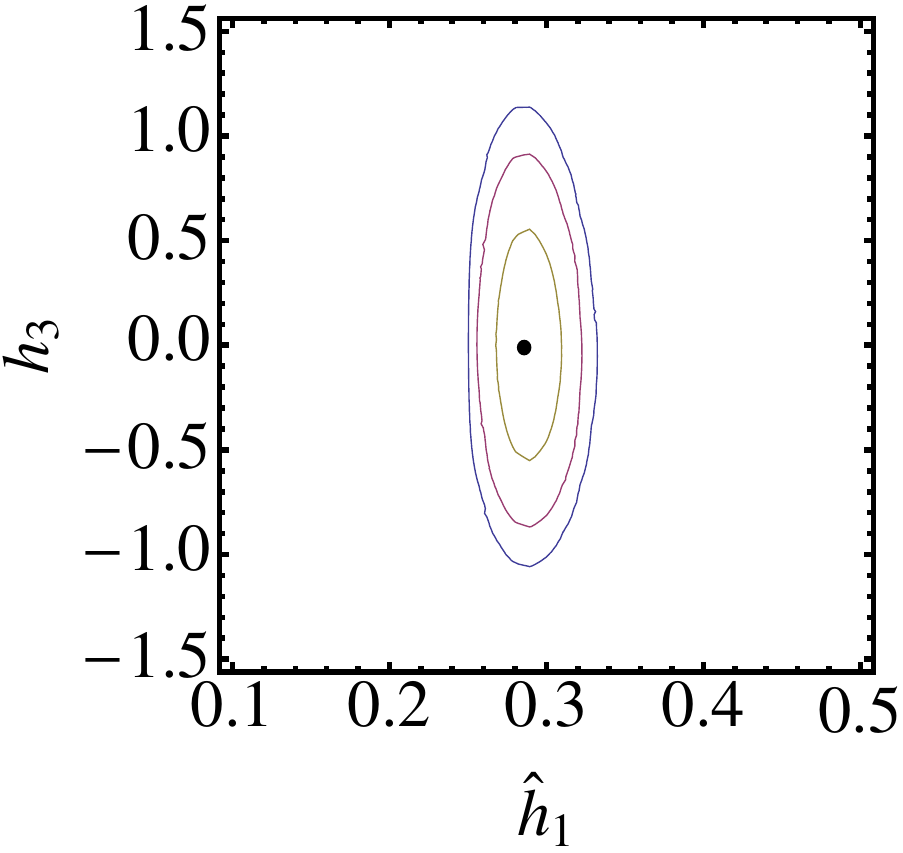}

\includegraphics[width=6.6cm]{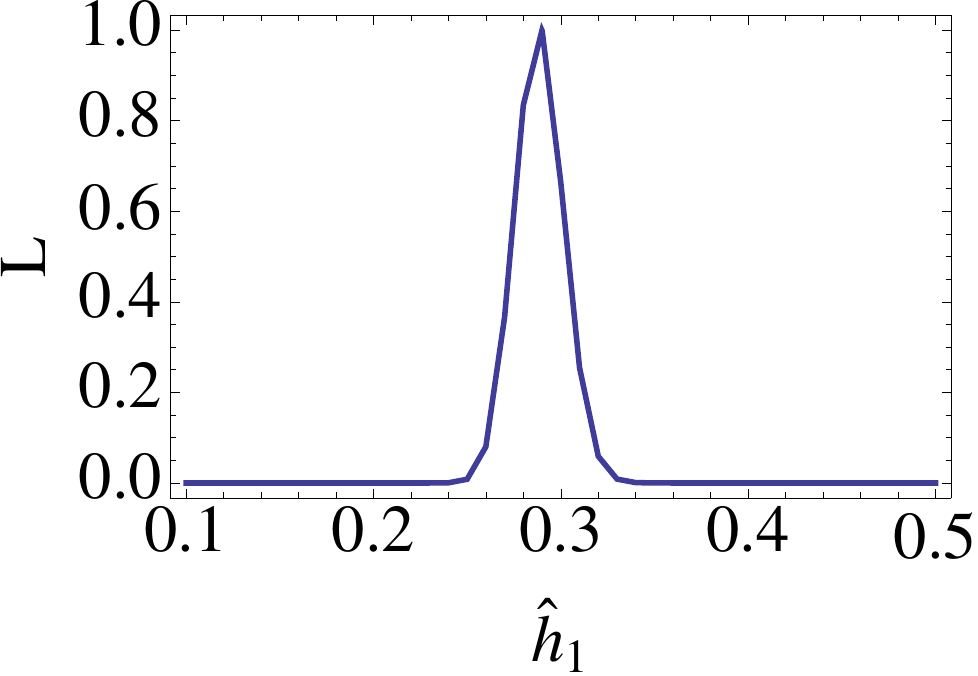}\includegraphics[width=6.6cm]{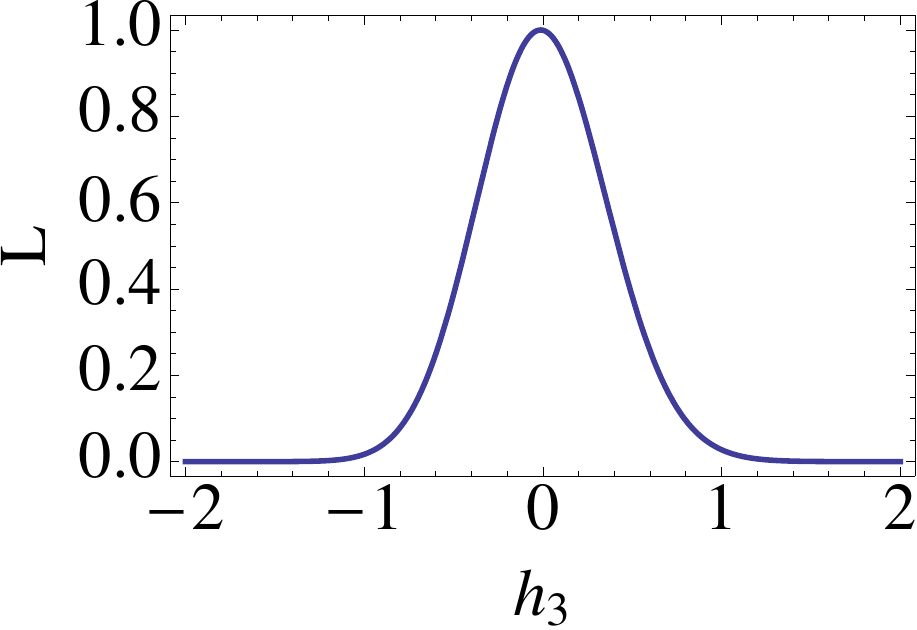}

\protect\protect\protect\caption{$\emph{Top panel }$: 1$\sigma$, 2$\sigma$ and 3$\sigma$ confidence-level
contours for the 2-dimensional forecast posterior on the parameters
$\{\hat{h}_{1},$$h_{3}\}$ marginalizing over $\sigma_{8}$ (fifth
case). $\emph{Left bottom panel }$\emph{: }forecast posterior for
$\hat{h}_{1}$ marginalized over $h_{3}$. $\emph{Right bottom panel }$:
forecast posterior for $h_{3}$ marginalized over $\hat{h}_{1}$.}

\label{Fig7} 
\end{figure}

\begin{figure}
\includegraphics[width=6cm]{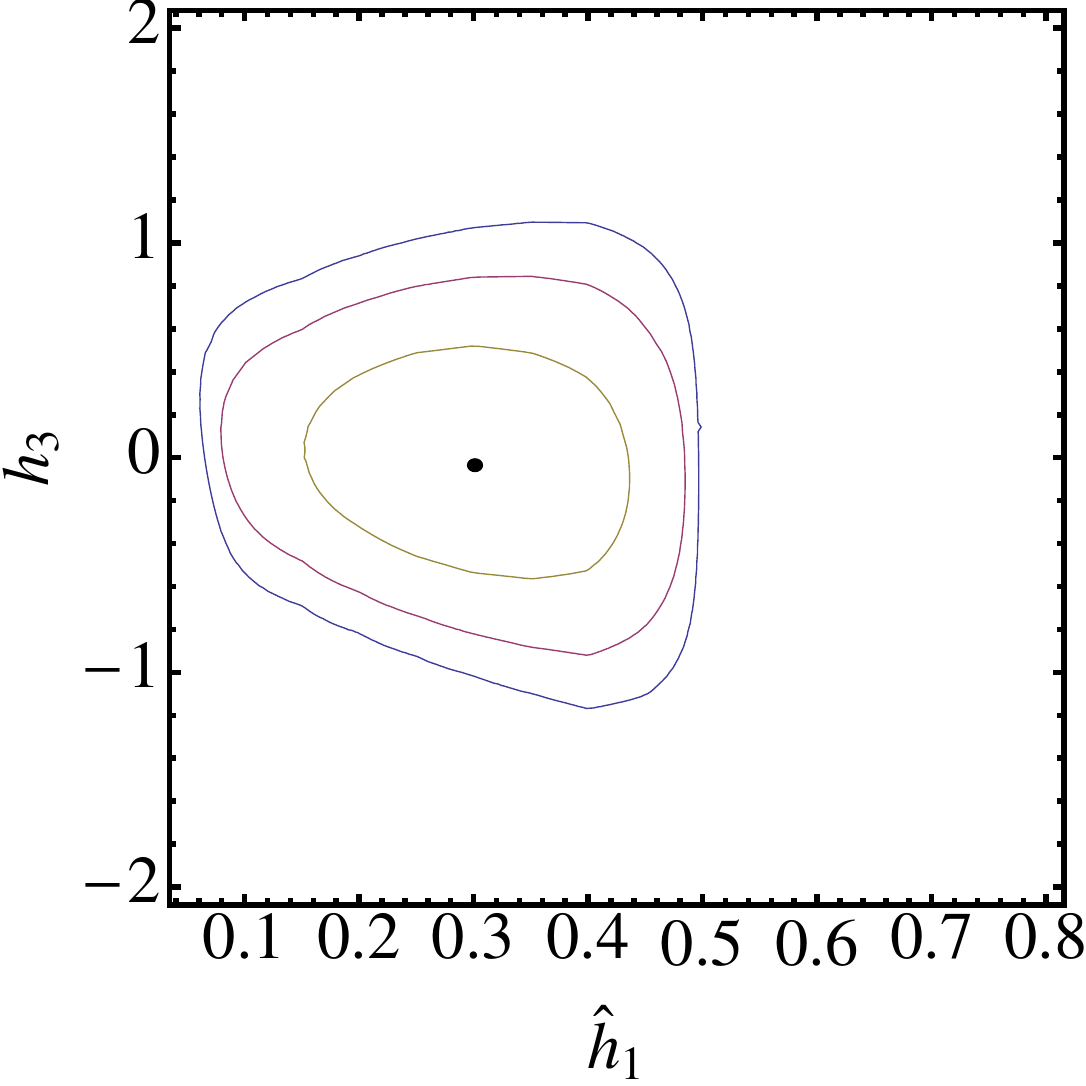}\includegraphics[width=6.6cm]{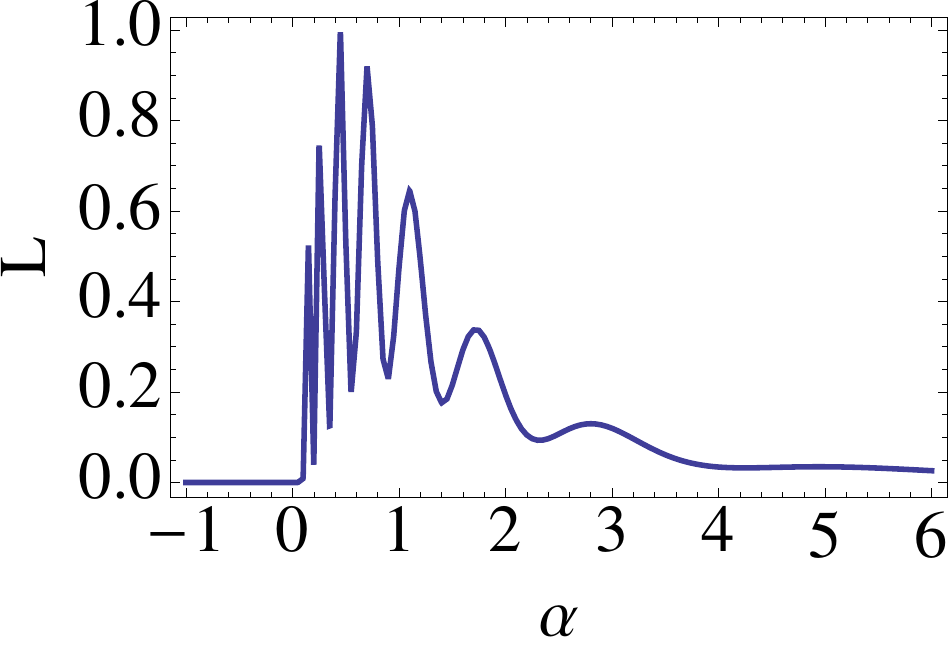}

\includegraphics[width=6.6cm]{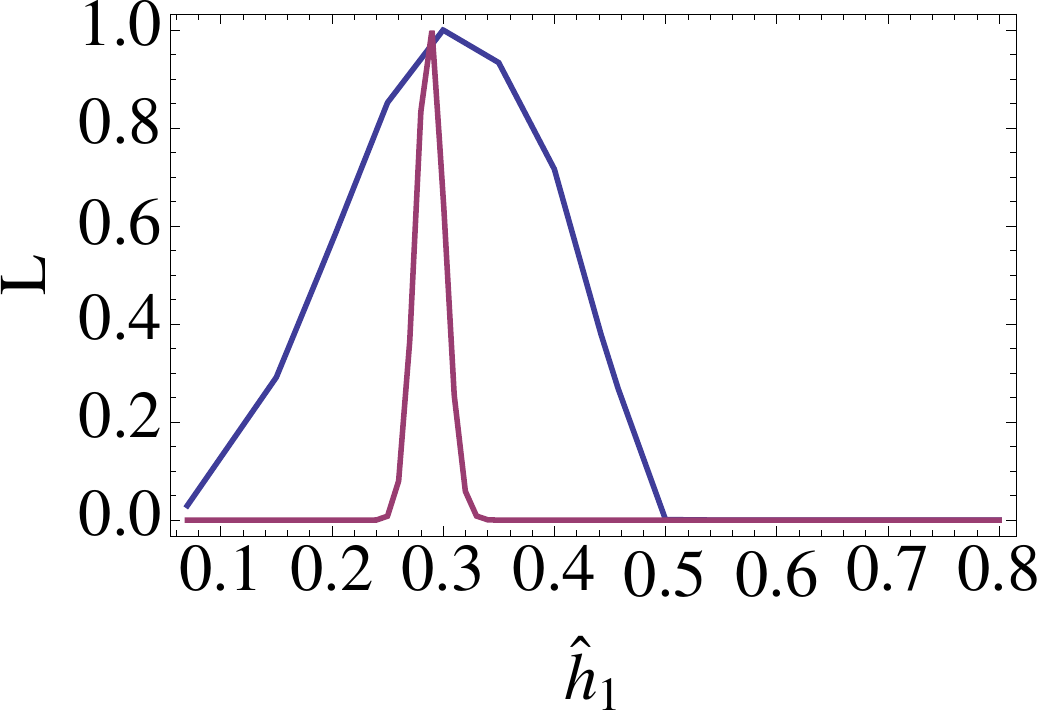}\includegraphics[width=6.6cm]{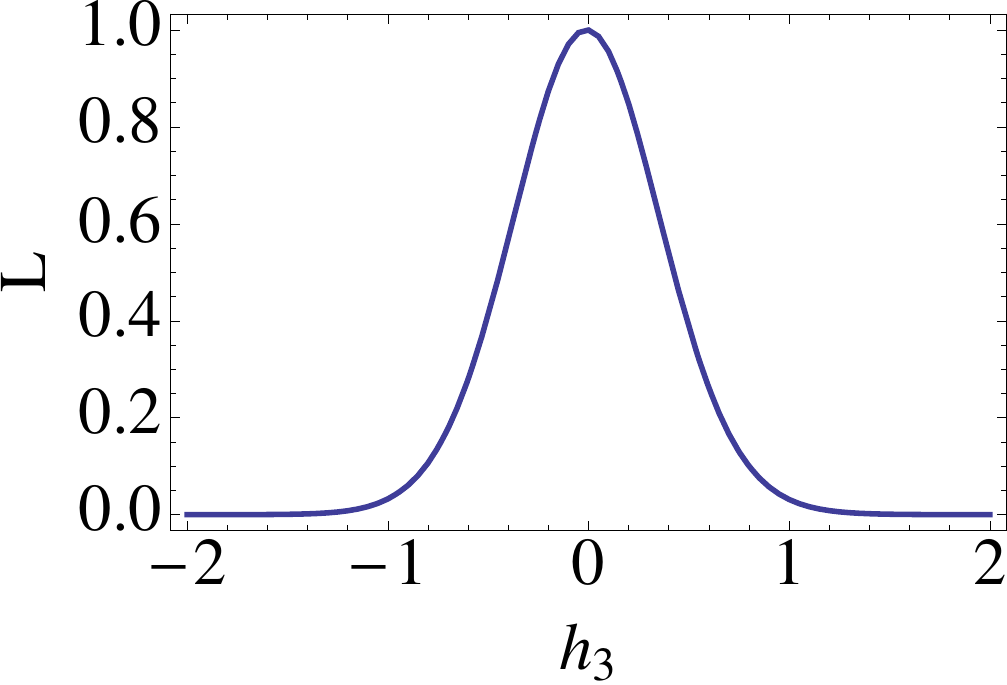}

\protect\protect\protect\caption{$\emph{Top panel }$: 1$\sigma$, 2$\sigma$ and 3$\sigma$ confidence-level
contours for the 2-dimensional\emph{ }forecast posterior on the parameters
$\{\hat{h}_{1},$$h_{3}\}$ marginalizing on $\{\sigma_{8}$, $\alpha\}$
(sixth case). $Left\emph{ bottom panel }$: forecast posterior marginalized
on $\{h_{3},\alpha\}$ varying the initial conditions (blue line)
in comparison with the fifth case (red line). $\emph{Right bottom panel }$:
forecast posterior marginalized on $\{\hat{h}_{1},\alpha\}$}

\label{Fig8} 
\end{figure}

The model now contains three parameters: $\{\hat{h}_{1,}h_{3},\alpha\}$.
Note that in principle one should take a different $\alpha$ for every
$k$ but for simplicity we assume that $\alpha$ is $k$-independent
in our range. As in the previous cases, here we analyze first the
case in which $\alpha=1$ (this is our \emph{fifth case}) and the
case in which we will vary this parameter (\emph{sixth case}). We
numerically solve Eq. (\ref{delta}) inserting now the value of $k$
corresponding to the central $k$-bin values for every redshift bin
and then we construct the $\sigma_{8}$-marginalized three dimensional
forecasted posterior by following the same procedure described in
section III. The results are reported in Table \ref{tab8} and in
Figs. (\ref{Fig7},\ref{Fig8}). The error on $\hat{h}_{1}$ increase
from roughly 0.02 to 0.10 when marginalizing over the initial condition.
In contrast, the error on the scale $h_{3}$ remain practically unchanged,
since we assume $k$-independent initial conditions.

\begin{table}
\begin{tabular}{|c|c|c|c|}
\hline 
$\bar{{z}}$  & $k_{min}-k_{1}$  & $k_{1}-k_{2}$  & $k_{2}-k_{max}$ \tabularnewline
\hline 
\hline 
0.6  & \multicolumn{1}{c|}{0.007-0.022} & 0.022-0.063  & 0.063-0.180 \tabularnewline
\hline 
0.8  & 0.007-0.023  & 0.023-0.071  & 0.071-0.215 \tabularnewline
\hline 
1.0  & 0.007-0.024  & 0.024-0.078  & 0.078-0.249 \tabularnewline
\hline 
1.2  & 0.007-0.026  & 0.026-0.086  & 0.086-0.287\tabularnewline
\hline 
1.4  & 0.007-0.027  & 0.027-0.094  & 0.094-0.329 \tabularnewline
\hline 
1.8  & 0.007-0.029  & 0.029-0.112  & 0.112-0.426 \tabularnewline
\hline 
\end{tabular}

\protect\protect\protect\protect\caption{Ranges of the $k$-bins for every redshift bin centered at $\bar{z}$,
in units of $(h/Mpc)$ (from \cite{Amendola:2013qna}).}

\label{tab4} 
\end{table}

\begin{table}
\begin{tabular}{|c|c|c|c|c|}
\hline 
$\bar{{z}}$  & $i$  & $f\sigma_{8}(z)$  & $\Delta$$f\sigma_{8}(z)$  & $\Delta f\sigma_{8}(z)\%$\tabularnewline
\hline 
\hline 
\multirow{3}{*}{0.6 } & 1  & \multirow{3}{*}{0.469 } & 0.07  & 15\tabularnewline
\cline{2-2} \cline{4-5} 
 & 2  &  & 0.017  & 3.6\tabularnewline
\cline{2-2} \cline{4-5} 
 & 3  &  & 0.0097  & 2.1\tabularnewline
\hline 
\multirow{3}{*}{0.8 } & 1  & \multirow{3}{*}{0.457 } & 0.05  & 11\tabularnewline
\cline{2-2} \cline{4-5} 
 & 2  &  & 0.012  & 2.6\tabularnewline
\cline{2-2} \cline{4-5} 
 & 3  &  & 0.0074  & 1.6\tabularnewline
\hline 
\multirow{3}{*}{1.0 } & 1  & \multirow{3}{*}{0.438 } & 0.039  & 8.9\tabularnewline
\cline{2-2} \cline{4-5} 
 & 2  &  & 0.0089  & 2\tabularnewline
\cline{2-2} \cline{4-5} 
 & 3  &  & 0.0062  & 1.4\tabularnewline
\hline 
\multirow{3}{*}{1.2 } & 1  & \multirow{3}{*}{0.417 } & 0.032  & 7.7\tabularnewline
\cline{2-2} \cline{4-5} 
 & 2  &  & 0.0072  & 1.7\tabularnewline
\cline{2-2} \cline{4-5} 
 & 3  &  & 0.0055  & 1.3\tabularnewline
\hline 
\multirow{3}{*}{1.4 } & 1  & \multirow{3}{*}{0.396 } & 0.028  & 7\tabularnewline
\cline{2-2} \cline{4-5} 
 & 2  &  & 0.0065  & 1.6\tabularnewline
\cline{2-2} \cline{4-5} 
 & 3  &  & 0.0057  & 1.4\tabularnewline
\hline 
\multirow{3}{*}{1.8 } & 1  & \multirow{3}{*}{0.354 } & 0.015  & 4.3\tabularnewline
\cline{2-2} \cline{4-5} 
 & 2  &  & 0.0047  & 1.3\tabularnewline
\cline{2-2} \cline{4-5} 
 & 3  &  & 0.0061  & 1.7\tabularnewline
\hline 
\end{tabular}

\protect\protect\caption{Fiducial values and relative errors for $f\sigma_{8}$ data at every
redshift $\bar{{z}}$ and every $k$-bin (labeled with the index $i$).}

\label{tab7} 
\end{table}

\begin{table}
\begin{tabular}{|c|c|c|c|c|c|c|c|c|c|c|}
\hline 
\multirow{2}{*}{} & \multirow{2}{*}{case } & $\alpha$  & \multirow{2}{*}{$\Delta\alpha$ (95\%) } & \multirow{2}{*}{$\Delta\alpha$ (68\%) } & $\hat{h}_{1}$  & \multirow{2}{*}{$\Delta\hat{h}_{1}$ (95\%) } & \multirow{2}{*}{$\Delta\hat{h}_{1}$ (68\%) } &  $h_{3}$  & \multirow{2}{*}{$\Delta h_{3}$ (95\%) } & \multirow{2}{*}{$\Delta h_{3}$ (68\%)}\tabularnewline
 &  & (best fit) &  &  & (best fit) &  &  & (best fit) &  & \tabularnewline
\hline 
\hline 
\multirow{2}{*}{Horndeski } & \multirow{2}{*}{V } & \multirow{2}{*}{1 } & \multirow{2}{*}{- } & \multirow{2}{*}{-} & \multirow{2}{*}{0.3 } & \multirow{2}{*}{{[}0.26, 0.32{]} } & \multirow{2}{*}{{[}0.27, 0.32{]} } & \multirow{2}{*}{0 } & \multirow{2}{*}{{[}-0.70, 0.72{]} } & \multirow{2}{*}{{[}-0.37, 0.35{]}}\tabularnewline
 &  &  &  &  &  &  &  &  &  & \tabularnewline
\hline 
\multirow{2}{*}{Horndeski} & \multirow{2}{*}{VI } & \multirow{2}{*}{0.85 } & \multirow{2}{*}{{[}0.10, 2.2{]} } & \multirow{2}{*}{{[}0.22, 1.9{]}} & \multirow{2}{*}{0.3 } & \multirow{2}{*}{{[}0.097, 0.44{]} } & \multirow{2}{*}{{[}0.17, 0.40{]} } & \multirow{2}{*}{0 } & \multirow{2}{*}{{[}-0.72, 0.73{]}} & \multirow{2}{*}{{[}-0.36, 0.36{]}}\tabularnewline
 &  &  &  &  &  &  &  &  &  & \tabularnewline
\hline 
\end{tabular}

\protect\protect\protect\protect\caption{Best fit and errors on $\hat{h_{1}}$, $h_{3}$ in the Horndeski case
by fixing $h_{5}=0$.}

\label{tab8} 
\end{table}

\section{A cosmological exclusion plot}

Here we wish to continue the analysis by obtaining an exclusion plot,
i.e. the region of parameter space that a future Euclid-like redshift
survey can achieve. This is obtained by repeating the procedure of
the previous section obtaining the errors on $\hat{h}_{1},h_{3}$
for every possible $h_{5}$ (rather than fixing $h_{5}$ to the standard
value). The region outside the errors is therefore the region that
an Euclid-like experiment will be able to rule out.

The form of $Y$ in Eq. (\ref{Y1}) produced in a Horndeski model
represents a Yukawa-like gravitational potential in real space. By
Fourier anti-transforming Eq. (\ref{Y}) with a point source of mass
$M$ one obtains in fact 
\begin{equation}
\Psi(r)=-\frac{G_{0}M}{r}h_{1}\left(1+Qe^{-r/\lambda}\right)
\end{equation}
where $h_{5}=(1+Q)\lambda^{2}$ and $h_{3}=\lambda^{2}$ (notice that
here again $Mh_{1}$ is the observable, not $h_{1}$ alone). Here
$G_{0}$ is the gravitational constant one would measure in laboratory
where, as already mentioned, the effects of the modification of gravity
are assumed to be screened %
\footnote{Notice that although one could define a new gravitational ``constant''
$G_{\mathrm{eff}}=G_{0}h_{1}(1+Qe^{-r/\lambda})$ in the potential,
one should use a different definition, namely $G_{\mathrm{eff}}^{F}=G_{0}h_{1}(1+Qe^{-r/\lambda}(1+r/\lambda))$,
in the force. This is why we prefer to use a different notation, i.e.
$Y$.%
}.

\begin{figure}
\includegraphics[width=0.6\textwidth]{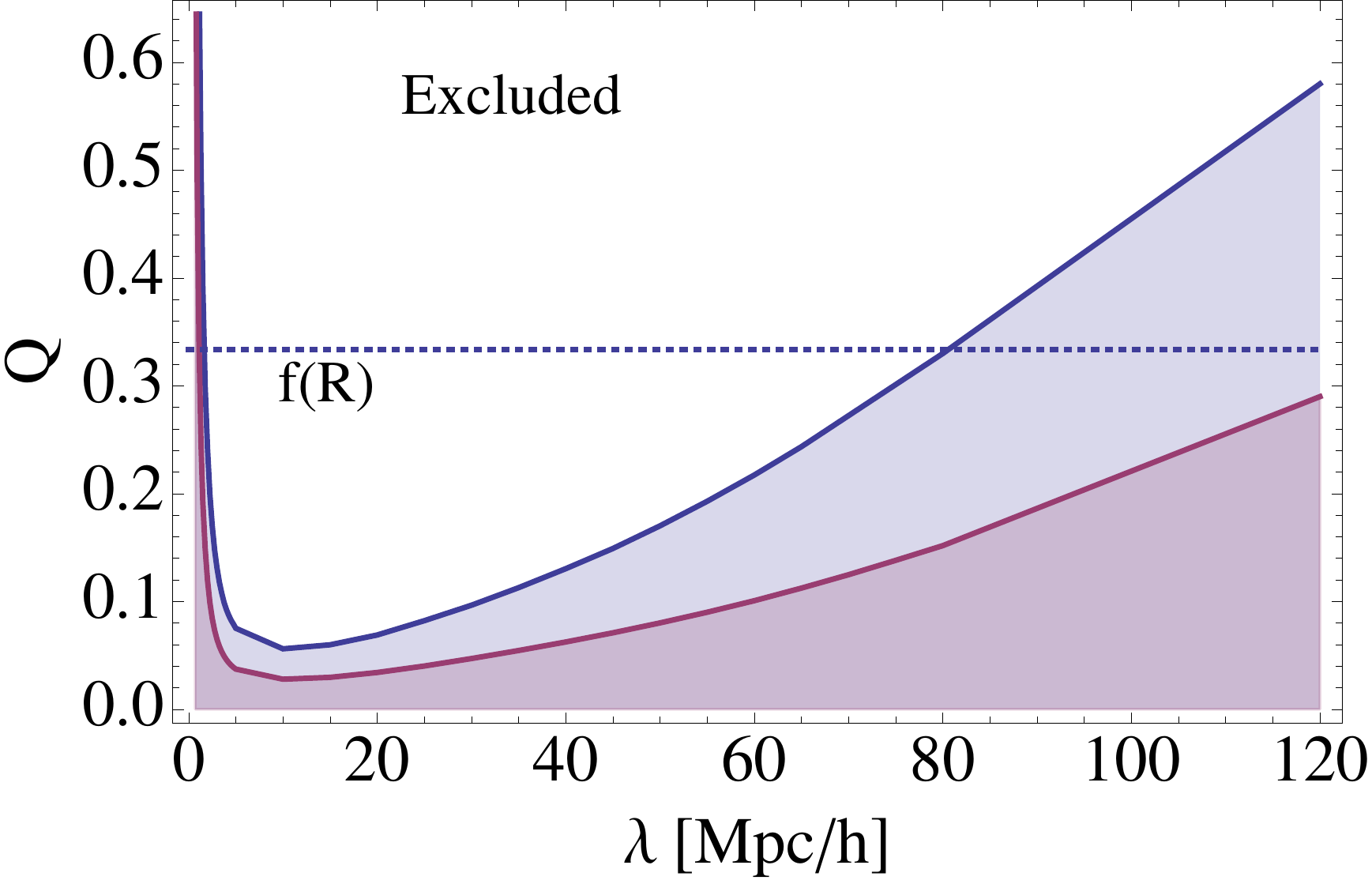}

\caption{\label{fig:Forecast-of-a}Forecast of a cosmological exclusion plot
for a Euclid-like survey, marginalizing over $\sigma_{8},\alpha$
and $\hat{h_{1}}$. Here $Q$ is the dimensionless strength of the
Yukawa interaction while $\lambda$, in Mpc$/h$, is the interaction
range. The darker region is the 68\% c.l. region, the lighter one
is the 95\%. c.l. region. The dotted line marks the value of $Q$
in $f(R)$ models.}
\end{figure}

Thus, instead of $h_{3,5}$, we can use the strength $Q$ and range
$\lambda$ of the Yukawa term as modified-gravity parameters, marginalizing
over $\hat{h}_{1}=\Omega_{m,0}h_{1}$ and, as before, also over $\sigma_{8}$
and $\alpha$. As previously, we assume $Q,\lambda$ to be constant
in the observed range. These parameters are the cosmological analog
of the parameters employed in laboratory experiments to test deviations
from Newtonian gravity, see e.g. \cite{2007PhRvL..98b1101K}. Using
the same specifications of the previous section, we show in Fig. (\ref{fig:Forecast-of-a})
the region that a Euclid-like experiment is able to exclude. Clearly,
for very small $\lambda$ the strength $Q$ is unconstrained; moreover,
for very large interaction ranges (much larger than the observed scales),
the strength becomes degenerate with $h_{1}$ and therefore again
weakly constrained. In the intermediate region around 10 Mpc$/h$
the strength can be confined to within 0.03 (0.06) at 68\% (95\%)
c.l., i.e. 3\% (6\%) of the Newtonian gravitational strength. This
limit is of course much weaker than local gravity bounds, which are
below $10^{-4}$, but it applies to scales and epochs unreachable
with other means. The results will not change much if we do not marginalize
over initial conditions, just as it happened for $h_{3}$ in the previous
section.

For comparison, the strength $Q$ in the case of $f(R)$ models is
1/3 (see e.g. \cite{DeFelice_Tsujikawa_2010}), while the range is
\begin{equation}
\lambda_{f(R)}=M_{f(R)}^{-1}=\sqrt{\frac{3f_{,RR}}{f_{,R}}}
\end{equation}
where the subscripts denote the derivative with respecto to $R$ of
the Lagrangian $f(R)$ (in this notation $f(R)$ includes the Einstein-Hilbert
term). From Fig. (\ref{fig:Forecast-of-a}) one can see that all the
models with $2\lesssim\lambda_{f(R)}\lesssim80$ Mpc$/h$ could be
ruled out at 95\% c.l. for $Q=1/3$. Conversely, assuming $f_{,R}\approx1$
as needed by local gravity constraints and by a background close to
$\Lambda$CDM, a Euclid-like survey will be able to set a lower and
an upper limit to $f_{,RR}$: 
\begin{eqnarray}
f_{,RR} & < & 1\cdot10^{-7}H_{0}^{-2}\,,\quad\mathrm{or}\\
f_{,RR} & > & 2\cdot10^{-4}H_{0}^{-2}
\end{eqnarray}
In keeping with our analysis, we are assuming here $\lambda_{f(R)}$
constant; in general however it will be a function of time so these
limits should refer to the epoch of observation. In some popular models
of $f(R)$ one has $f_{,RR}\approx10^{-3}H_{0}^{-2}$ at $z\approx1$
(see e.g. \cite{2012arXiv1206.1642J}), corresponding to $\lambda_{f(R)}\approx100-200$
Mpc$/h$, a value that could be marginally detected at 68\% c.l. by
our forecasts.

Notice however that in $f(R)$ models the overall factor here denoted
as $\hat{h}_{1}$ corresponds to $\Omega_{m,0}/f_{'R}\approx\Omega_{m,0}$.
The existence of a lower limit to $f_{,RR}$ is due to the marginalization
over the unknown $\Omega_{m,0}$. In specific models of $f(R)$ the
present matter density $\Omega_{m,0}$ can be estimated through background
or large-scale structure measurements. In this case the lower limit
would be removed and any $\lambda_{f(R)}$ larger than a few Megaparsec
would be detected. The application of the results of this paper to
specific models of modified gravity is left to future work.

\section{Conclusions}

In this paper we investigated the current and future bounds on the
modified gravity parameter $Y$ (or $G_{\mathrm{eff}}$) that quantifies
the deviation from the standard Poisson equation. We have assumed
$Y$ to be constant in time and space when using current data or with
a Horndeski behavior when forecasting future results. Contrary to
other similar analyses, we tried to weaken the model-dependency by
marginalizing over the present power spectrum normalization $\sigma_{8}$
and over the initial growth rate for the matter density contrast equation,
since they both are unknown unless one assumes a specific model, e.g.
$\Lambda$CDM. We also take into account the fact that $\Omega_{m,0}$
is not a directly observable quantity and absorb it into the definition
of $Y$.

We find, not unexpectedly, that the current growth rate data $f\sigma_{8}(z)$
from redshift distortion are insufficient to constrain the product
$\hat{Y}=\Omega_{m,0}Y$ to better than an order of 100\% error (see
Table \ref{tab:Summary-of-results-1}, fourth case), due to the degeneracy
with $\sigma_{8}$ and the initial condition. Using instead forecasts
of a Euclid-like experiment, we find that the relative error on $\Omega_{m,0}Y$
reduce to roughly 30\% at 68\% c.l. (see Table \ref{tab3}, fourth
case). A similar error can be obtained on $\hat{h}_{1}=\Omega_{m,0}h_{1}$
when using the Horndeski prescription (see Table \ref{tab8}, sixth
case). The effect of the lack of knowledge of the initial conditions
can be easily grasped by noting that the uncertainty on $\hat{Y}$
increases from $\Delta\hat{Y}\approx0.01$ when $\alpha=1$ to $\Delta\hat{Y}\approx0.08$
when $\alpha$ is marginalized over (Table \ref{tab:Summary-of-results-1}),
i.e. from a few percent to 30\%. Same broadening of the uncertainty
occurs for $\sigma_{8}$.

Finally, we obtain a forecast of a cosmological exclusion plot on
the Yukawa strength $Q$ and range $\lambda$ parameters (Fig. \ref{fig:Forecast-of-a}).
This complements, on cosmological scales, the laboratory exclusion
plots on deviations from standard gravity. We find that with a Euclid-like
experiment the strength $Q$ can be confined to within 3\%(6\%) of
the Newtonian gravity at 68\%(95\%) if the interaction range is around
10 Megaparsecs. For much larger and much smaller ranges the constraint
gradually vanishes. Applying these results to $f(R)$ models we forecast
an upper limit to $f_{,RR}$ at $z\approx1$ of the order of $10^{-7}H_{0}^{-2}$,
corresponding to a Yukawa range smaller than 2 Mpc$/h$ roughly, and
a lower limit of $2\cdot10^{-4}H_{0}^{-2}$, corresponding to scales
larger than $80$Mpc$/h$ (at 95\% c.l.).

The main conclusion of this paper is that $Y$ can be only weakly
constrained by the next decade redshift surveys if one takes into
account the degeneracy with $\sigma_{8},\Omega_{m,0}$ and initial
conditions. Even weaker constraints would have been obtained had we
taken $Y$ to be time dependent. Only by considering specific models
can one hope to produce stringent constraints on modified gravity
through its effect on linear matter perturbation growth. This seems
to indicate that the other modified gravity linear perturbation parameter,
the anisotropic stress $\eta$, which requires a combination of weak
lensing and clustering, is a more robust and powerful way to quantify
the deviation from standard gravity.

\begin{acknowledgments}
L.A. acknowledges support from DFG through the project TRR33 ``The
Dark Universe''. We thank Alejandro Guarnizo-Trilleras and Adrian
Vollmer for help with the forecasts and Guillermo Ballesteros, Emilio
Bellini, Valerio Marra and Valeria Pettorino for useful discussions.
L.T. thanks the Institute of Theoretical Physics at the University
of Heidelberg for the hospitality.
\end{acknowledgments}
\bibliography{observables,amendola,massive-gravity,growth-rate}

\end{document}